\documentclass{article}

\usepackage{microtype}
\usepackage{graphicx}
\usepackage{subcaption}
\usepackage{booktabs} 
\usepackage{tikz}
\usepackage{xcolor}
\usetikzlibrary{arrows.meta, positioning, calc}

\definecolor{amethyst}{HTML}{E8E0F0}
\definecolor{amethystB}{HTML}{7B5EA7}
\definecolor{sapphire}{HTML}{D6E4F0}
\definecolor{sapphireB}{HTML}{3A6FA0}
\definecolor{emerald}{HTML}{D4EDDA}
\definecolor{emeraldB}{HTML}{2D8659}
\definecolor{ruby}{HTML}{F5DCDC}
\definecolor{rubyB}{HTML}{B34B4B}
\definecolor{topaz}{HTML}{FDF0D5}
\definecolor{topazB}{HTML}{C08B30}
\usepackage{stfloats}
\usepackage{placeins}
\usepackage[ruled,vlined,linesnumbered]{algorithm2e}

\newcommand{\regime}[3]{\textbf{#1} $[\hat{H}{#2},\, E{#3}]$\textbf{.}}

\usepackage{hyperref}
\usepackage[accepted]{icml2026}

\usepackage{amsmath}
\usepackage{amssymb}
\usepackage{mathtools}
\usepackage{amsthm}
\usepackage{bm}  

\usepackage{tikz}
\usetikzlibrary{arrows.meta, positioning, shapes.geometric, calc, decorations.pathreplacing, calligraphy}

\usepackage{pgfplots}
\pgfplotsset{compat=1.18}
\usepackage{pgf-pie}
\usepgfplotslibrary{colorbrewer}  

\usepackage[most]{tcolorbox}
\newtcolorbox{classbox}{
  colback=black!5,      
  colframe=black!5,     
  boxrule=0pt,          
  arc=2pt,              
  left=6pt, right=6pt, top=4pt, bottom=4pt,
  breakable             
}

\definecolor{srcblue}{HTML}{1565C0}
\definecolor{srcbluebg}{HTML}{E3F2FD}
\definecolor{metricgreen}{HTML}{2E7D32}
\definecolor{metricgreenbg}{HTML}{E8F5E9}
\definecolor{phasepurple}{HTML}{6A1B9A}
\definecolor{phasepurplebg}{HTML}{F3E5F5}
\definecolor{darkgray}{HTML}{424242}
\definecolor{abstainred}{HTML}{C62828}
\definecolor{abstainredbg}{HTML}{FFEBEE}
\definecolor{tiergold}{HTML}{F57F17}
\definecolor{tiergoldbg}{HTML}{FFF8E1}
\definecolor{dynblue}{HTML}{0D47A1}
\definecolor{dynbluebg}{HTML}{E3F2FD}

\definecolor{bt1}{HTML}{E8F5E9}
\definecolor{bt2}{HTML}{FFFDE7}
\definecolor{bt3}{HTML}{FFF3E0}
\definecolor{bt4}{HTML}{FFEBEE}
\definecolor{bt5}{HTML}{F3E5F5}

\usepackage{colortbl}
\usepackage{tabularx}
\usepackage{float}
\usepackage{makecell}

\usepackage[utf8]{inputenc}
\usepackage{enumitem}

\usepackage[capitalize,noabbrev]{cleveref}

\theoremstyle{plain}

\theoremstyle{definition}

\theoremstyle{remark}

\usepackage[textsize=tiny]{todonotes}

\icmltitlerunning{Classification Framework for AI Incident Monitoring}

\begin{document}

\twocolumn[
  \icmltitle{A Pragmatic Classification Framework for AI Incident Monitoring}



  \icmlsetsymbol{equal}{*}

    \begin{icmlauthorlist}
    \icmlauthor{Isaak Mengesha}{arcadia,uva}
    \icmlauthor{Branwen Owen}{arcadia}
    \icmlauthor{Charlie Collins}{arcadia}
    \icmlauthor{Tina Wong}{arcadia}\hspace{20em}
    \icmlauthor{Simon Mylius}{mitft}
    \icmlauthor{Peter Slattery}{mitft}
    \icmlauthor{Sean McGregor}{raic}

  \end{icmlauthorlist}

    \icmlaffiliation{arcadia}{Arcadia Impact AI Governance Taskforce}
    \icmlaffiliation{uva}{Computational Science Lab, University of Amsterdam, Noord Holland, Netherlands}
    \icmlaffiliation{mitft}{MIT FutureTech, Massachusetts Institute of Technology, Cambridge, MA 02139, USA}
    \icmlaffiliation{raic}{Responsible AI Collaborative, Los Angeles, CA, USA}
  \icmlcorrespondingauthor{Isaak Mengesha}{isaak.mengesha@arcadiaimpact.org}

  \icmlkeywords{AI governance, incident monitoring, incident classification}

  \vskip 0.3in
]



\printAffiliationsAndNotice{}  

\begin{abstract}


Incident monitoring can drive safety improvements in high-reliability industries and population-scale technologies, but remains underdeveloped in AI governance. In the absence of systematic incident reporting, monitoring must draw on all publicly available sources and account for their biases and coverage limitations. Raw incident counts from any such source conflate reporting propensity, scale of AI system deployment (``exposure''), and frequency of harm per unit exposure. Drawing on epidemiological principles, we propose a methodological framework in three parts: a monitoring question that defines the scope of analysis; a procedure for estimating harm and exposure trends and calibrating confidence to the data available and assumptions made; and a classification scheme that maps trend estimates to actionable governance categories (Escalating, Mitigating, Concentrating, Receding, or Unclassifiable). Through case studies, we demonstrate insight despite data constraints and provide a proof of concept for AI incident monitoring as a practical governance tool.

\end{abstract}

\section{Introduction}\label{sec:introduction}

Since the launch of ChatGPT in 2022, general-purpose artificial intelligence (AI) systems have reached billions of users, while AI systems in general have been increasingly deployed in high-stakes settings, including healthcare and critical infrastructure \cite{bengio2026iaisr}. Despite significant technical advances, reliability challenges persist, and malfunctions, systemic disruptions and malicious use continue to cause documented harm \cite{bengio2026iaisr, jeanmaire2025incidents}. Systematic incident monitoring has driven safety improvements across other high-reliability industries and population-scale technologies, including aviation, nuclear power, pharmacovigilance, and healthcare, but AI currently lacks both the institutional infrastructure to support consistent reporting and the analytical frameworks to interpret what is reported \cite{paeth2025risk, wei2025designing}. This paper responds to the latter gap by proposing a simple methodology for estimating and comparing AI-related harm and exposure-to-harm between two time periods using publicly available data.

\begin{table}[H]
\centering
\footnotesize
\setlength{\tabcolsep}{4pt}
\renewcommand{\arraystretch}{1.15}
\begin{tabular}{@{}p{0.04\columnwidth}p{0.40\columnwidth}p{0.46\columnwidth}@{}}
\toprule
& \textbf{Problem} & \textbf{Framework response} \\
\midrule
P1 & Complexity and lack of standardization confounds analytical consistency & Structured monitoring question (\textsc{Sort}; \S\ref{sec:sort})\\
P2 & More harmful AI systems indistinguishable from more deployed AI systems & Separate exposure and harm estimates (\S\ref{sec:estimate}) \\
P3 & Sparse or missing data treated as absence of harm & Evidence tiers; proxy measures; confidence statements; principled abstention (\S\ref{sec:classification}) \\
P4 & Absolute values sensitive to reporting bias and uncertainty & Trend estimates more robust to stable biases and uncertainties (\S\ref{sec:uc})\\
& & \\
\bottomrule
\end{tabular}
\caption{\textbf{Four problems with raw incident counts,} and corresponding framework responses.}
\label{tab:problems}
\end{table}
The existing mandatory incident reporting obligations for AI remain narrow in scope and fragmented across jurisdictions. The EU AI Act (Articles 73 and 55) \cite{euaiact2024} requires providers and deployers of ``high-risk'' AI systems, and providers (but not deployers) of ``systemic-risk'' general-purpose AI models, to report serious incidents to national authorities and the EU AI Office, respectively. These reports trigger individual investigations, but reach different parts of the EU system with no mandate to consolidate or publicly share them, and the definition of a ``serious incident'' excludes many harms that any common-sense reading would consider serious. California's Transparency in Frontier Artificial Intelligence Act, TFAIA (SB-53) \cite{tfaia2025} and, from 2027, New York's Responsible AI Safety and Education (RAISE) Act \cite{raiseact2025}, are narrower still, applying only to a handful of frontier AI developers and capturing only catastrophic-scale harms, security breaches, and loss-of-control events, with no public disclosure requirement. As of May 2026, no other jurisdiction mandates AI-specific incident reporting or encourages voluntary reporting, leaving the vast majority of global AI deployments outside any incident reporting framework.

\begin{figure*}[b!]
  \centering
  \resizebox{0.98\textwidth}{!}{\input{Figures/Overview}}
  \caption{\textbf{The interpretive pipeline for AI incidents. } \textit{At top:} rising counts of monthly AI incidents and hazards (defined here as ``potential dangers") reported in the OECD AI Incidents and Hazards Monitor (AIM) conflate trends in media attention, deployment growth, and the frequency of harm per use of AI systems \cite{oecd2024aim}. \textit{Below:} our framework takes recorded incidents as inputs to structured monitoring questions, then estimates harm and exposure trends separately to classify each monitoring question into a trajectory category.}
  \label{fig:pipeline_v3}
\end{figure*}

Separately, the EU AI Act (Article 72) and the accompanying General-Purpose AI Code of Practice (Measure 3.5) \cite{euaiactcop2025} impose various post-market monitoring obligations on providers of high-risk AI systems and systemic-risk general-purpose AI models respectively. Both require the collection and analysis of relevant data through a range of internal and external methods, but neither instrument specifies an analytical methodology for doing so. As such, and despite the OECD developing standardized incident definitions \cite{oecd2024definitions, wei2025designing}, there is currently no agreement on how AI incident data should be analyzed or compared over time, and no validated methodologies for post-deployment AI monitoring \cite{rao2026challenges, whittlestone2021and}. This presents an opportunity to establish common approaches before national regimes fragment and solidify.

Demand for such methods is broader than the EU’s mandated post-market monitoring: in structured interviews with AI risk and safety practitioners---including government safety institutes, civil society, industry practitioners, and academic researchers---we found a consistent need for clarity on which harms are increasing in frequency or severity, and which are relevant to specific operational contexts. Our methodology is designed with this breadth of practitioner in mind.

In the absence of systematic reporting, we draw on the available public sources and acknowledge or account for their biases and coverage limitations to bridge the gap between what is recorded and what is actually happening. For example, news-sourced public databases (which may most closely resemble prospective systematic reporting databases) now catalog thousands of incidents involving AI systems \cite{oecd2024definitions, oecd2024aim, mcgregor2021aiid}. However, they currently capture only a fraction of all AI harms and, due to their sourcing, tend to over-represent acute, dramatic events  reported in English and under-represent chronic, diffuse, and systemic ones \cite{richards2025incidents, nixon2011slow, teo2025artificial} (see Appendix A). More fundamentally, raw incident counts from any monitoring source conflate at least three distinct factors: the propensity for incidents to be observed and reported; the scale of AI system deployment and use (``exposure''); and the frequency of harm per unit exposure \cite{paeth2025lessons, richards2025incidents}. Disaggregating these factors is a requirement of any monitoring regime to enable decision makers to understand and distinguish genuine risk escalation from artifacts of reporting, such as increased media attention \cite{paeth2025risk}.

In this respect, the field of epidemiology offers a useful comparison to AI. Like pathogens, AI systems interact with populations of unknown size, exposure is difficult to measure directly, and reporting is incomplete and uneven \cite{stein2024role, pratt2025documenting}. Epidemiological methods are designed for uncertainty: ``the art of epidemiologic reasoning is to draw sensible conclusions from imperfect data'' \cite{cdc2025fieldepi}. The core principles, including standardized case definitions and separation of exposure from harm rate \cite{mccarty1993ascertainment}, translate directly to the methodology we propose here. We draw on a variety of publicly available data sources with the aim of validating an analytical approach with broad applicability. News-sourced incident databases tend to be sparse on any specific topic but offer breadth across harm types and rich contextual detail, whereas domain-specific monitoring sources tend to be more representative but do not exist for all harm types.

This paper introduces the proposed methodological framework in three parts: a structured approach to defining an AI ``monitoring question''; procedures for estimating harm and exposure trends; and a classification process that maps combinations of trends to one of four broad trajectories with distinct governance implications. We demonstrate the methodology on a monitoring question on conversational AI and self-harm, and discuss its strengths, uses and limitations, finding that meaningful analysis is possible even with the imperfect data publicly available today.

\section{Framework}\label{sec:framework}

\begin{figure*}[b!]
  \centering
  \resizebox{0.98\textwidth}{!}{\begin{tikzpicture}[
    >={Stealth[length=4pt, width=3pt]},
    elbox/.style={
        rectangle, rounded corners=3pt, draw=#1B!60, fill=#1,
        minimum height=0.75cm, minimum width=2.2cm,
        align=center, font=\sffamily\small\bfseries,
        line width=0.5pt,
    },
    eletter/.style={
        circle, fill=#1B, text=white,
        minimum size=0.45cm, inner sep=0pt,
        font=\sffamily\small\bfseries,
    },
    qdesc/.style={font=\sffamily\scriptsize, text=black!60, align=left},
    arr/.style={->, line width=0.6pt, black!35},
    exbox/.style={
        rectangle, rounded corners=2pt, draw=#1B!40, fill=#1!50,
        minimum height=0.55cm, align=center,
        font=\sffamily\scriptsize, line width=0.3pt,
        text width=3.2cm,
    },
]

\def\xsep{3.8cm}

\node[eletter=ruby] (Sl) at (0, 0) {S};
\node[elbox=ruby, right=0.15cm of Sl] (Sb) {Subject};
\node[qdesc, below=2pt of Sb] {Who or what is at risk?};

\node[eletter=sapphire] (Ol) at (\xsep, 0) {O};
\node[elbox=sapphire, right=0.15cm of Ol] (Ob) {Opportunity};
\node[qdesc, below=2pt of Ob] {What creates the exposure?};

\node[eletter=emerald] (Rl) at (2*\xsep, 0) {R};
\node[elbox=emerald, right=0.15cm of Rl] (Rb) {Risk event};
\node[qdesc, below=2pt of Rb] {What specific harm?};

\node[eletter=topaz] (Tl) at (3*\xsep, 0) {T};
\node[elbox=topaz, right=0.15cm of Tl] (Tb) {Timeframe};
\node[qdesc, below=2pt of Tb] {Over what period?};

\draw[arr] (Sb.east) -- (Ol.west);
\draw[arr] (Ob.east) -- (Rl.west);
\draw[arr] (Rb.east) -- (Tl.west);

\node[rectangle, rounded corners=3pt, draw=black!20, fill=black!3,
    minimum height=0.7cm, font=\sffamily\small, text=black!70,
    inner xsep=10pt, line width=0.4pt]
    (pattern) at (0.5*3*\xsep + 0.5*2.2cm, -1.6)
    {``Among [\textcolor{rubyB}{\textbf{S}}] that [\textcolor{sapphireB}{\textbf{O}}], how many [\textcolor{emeraldB}{\textbf{R}}] per [\textcolor{topazB}{\textbf{T}}]?''};

\node[font=\sffamily\scriptsize\bfseries, text=black!50, anchor=west]
    (exlabel) at (-0.3, -2.6) {EXAMPLE};

\node[exbox=ruby, anchor=west] (exS) at (-0.3, -3.3)
    {Registered AV-capable\\vehicles};
\node[exbox=sapphire, right=0.25cm of exS] (exO)
    {Operating in self-driving\\mode on public roads};
\node[exbox=emerald, right=0.25cm of exO] (exR)
    {Cause injury or \\loss of life};
\node[exbox=topaz, right=0.25cm of exR] (exT)
    {Per million\\vehicle-miles};

\node[font=\sffamily\scriptsize\bfseries, text=rubyB] at (exS.north) [above=1pt] {S};
\node[font=\sffamily\scriptsize\bfseries, text=sapphireB] at (exO.north) [above=1pt] {O};
\node[font=\sffamily\scriptsize\bfseries, text=emeraldB] at (exR.north) [above=1pt] {R};
\node[font=\sffamily\scriptsize\bfseries, text=topazB] at (exT.north) [above=1pt] {T};

\end{tikzpicture}}
  \caption{\textbf{The SORT framework.} Structured monitoring questions reduce ambiguity in analysis due to ill-defined incident types.}
  \label{fig:sort}
\end{figure*}
\FloatBarrier
\paragraph{Why ask a monitoring question?}
A precisely defined monitoring question (MQ) helps ensure comparability, reproducibility, and internal consistency in the estimation and classification steps.

The problems in Table~\ref{tab:problems} arise due to the conflation of three steps: question definition, estimation, and interpretation. Our methodology takes each in turn: first, a precise monitoring question (MQ) is defined; second, trends in exposure and harm are estimated and confidence is calibrated to available evidence (\S\ref{sec:estimate}); and third, the resulting pair of trends is combined and classified into one of four trajectory categories (\S\ref{sec:classification}). 

\paragraph{Structuring the monitoring question.}\label{sec:sort}

Our SORT framework structures AI incident monitoring questions around four components: Subject (who or what is at risk), Opportunity (what exposes the subjects to the harm mechanism), Risk event (the countable harm), and Timeframe (the observation period), producing questions of the form ``Among [S] that [O], how many [R] per [T]?'' (See figure \ref{fig:sort} for a worked example.) We drew inspiration from the well-established PICO framework (Patient/Problem, Intervention, Comparison, and Outcome) in evidence-based medicine \cite{richardson1995pico}, to which SORT is similarly flexible: [S] and [O] can refer to people, systems, content, conversations or deployments, and can focus on the causes of harm, or victims of harm, allowing the same underlying harm to generate multiple valid MQs depending on the analyst's needs, while forcing analytical choices to be explicit rather than buried in unstated assumptions. [T] can be set to accommodate different harm dynamics and data availability. An interactive SORT tool is available to guide users in developing a precise MQ \footnote{Available as a \href{https://claude.ai/public/artifacts/9ec3813a-399c-48ea-96ed-19443d121b83}{Claude artifact} (see GitHub).}.


\FloatBarrier
\subsection{Estimation procedures} \label{sec:estimate}
Answering an MQ requires estimating two variables across consecutive time periods: the total harm associated with [R], and the exposure defined by [S] and [O]. We categorize methods for estimations into four tiers (Table~\ref{tab:tiers}) according to the strength of the available evidence and the corresponding confidence we can have in the results.

\begin{table}[H]
\centering
\caption{Tiers for harm and exposure estimation.}
\setlength{\tabcolsep}{3pt}
\small
\begin{tabular}{@{}llll@{}}
\toprule
Tier & Method & Sensitive to & Confidence \\
\midrule
1 & Direct measurement & Authoritative source  & High   \\
2 & Combine proxy measures & Proxy construction  & Medium \\
3 & Expert elicitation & Panel selection     & Low    \\
4 & Abstain             & ---                 & ---    \\
\bottomrule
\end{tabular}
\label{tab:tiers}
\end{table}

At Tier 1, the estimate is read directly from an authoritative source, such as vehicle crash filings, pharmacovigilance registries, or platform transparency reports. At Tier 2, no single source provides complete data, so the analyst constructs a \emph{point estimate} by combining partial sources and proxy measures. Estimates must be positive, by construction, and the distribution of possible value tends to be skewed to the right. For harm, public incident databases supply a hard \emph{lower bound}: the true harm cannot fall below what has already been recorded. At Tier 3, no publicly available quantitative sources support even a rough estimate, and domain experts are asked to bound a plausible range. Where the plausible range spans more than two orders of magnitude, or no expert consensus can be reached, the estimate is assigned to Tier 4, principled abstention. This is a valid finding in its own right: that current evidence cannot support even an order-of-magnitude estimate. Our use of structured qualitative certainty labels to communicate evidential strength where precise statistical uncertainty quantifications are not feasible follows established practice in climate science  and clinical guideline development \cite{dethier2023,guyatt2004} .

\begin{figure}[h]
    \centering
    \resizebox{\columnwidth}{!}{%
\definecolor{neutralBG}{HTML}{F1EFE8}
\definecolor{greenBG}{HTML}{E1F5EE}\definecolor{greenLine}{HTML}{1D9E75}\definecolor{greenTx}{HTML}{0F6E56}
\definecolor{blueBG}{HTML}{E6F1FB}\definecolor{blueLine}{HTML}{378ADD}\definecolor{blueTx}{HTML}{185FA5}
\definecolor{amberBG}{HTML}{FAEEDA}\definecolor{amberLine}{HTML}{EF9F27}\definecolor{amberTx}{HTML}{854F0B}
\definecolor{pinkBG}{HTML}{FBEAF0}\definecolor{pinkLine}{HTML}{D4537E}\definecolor{pinkTx}{HTML}{993556}
\definecolor{amyBG}{HTML}{EEEDFE}\definecolor{amyLine}{HTML}{7F77DD}\definecolor{amyTx}{HTML}{26215C}

\begin{tikzpicture}[
  >={Stealth[length=2.2mm]},
  arr/.style={->, line width=0.6pt, black!45},
  fork/.style={line width=0.6pt, black!45},
  darr/.style={->, line width=0.5pt, black!35, dashed},
  dec/.style={diamond, draw=black!30, fill=neutralBG, aspect=2.2,
    align=center, font=\sffamily\small\bfseries, text=black!85,
    inner sep=1pt, minimum width=2.3cm},
  proc/.style={rectangle, rounded corners=4pt, draw=amyLine, fill=amyBG,
    align=center, font=\sffamily\small\bfseries, text=amyTx,
    minimum width=2.2cm, minimum height=0.9cm},
  tier/.style={rectangle, rounded corners=5pt, align=center,
    font=\sffamily\footnotesize, minimum width=3.1cm, minimum height=0.95cm,
    line width=0.7pt},
  tHigh/.style={tier, draw=greenLine, fill=greenBG, text=greenTx},
  tMed/.style ={tier, draw=blueLine,  fill=blueBG,  text=blueTx},
  tLow/.style ={tier, draw=amberLine, fill=amberBG, text=amberTx},
  tAb/.style  ={tier, draw=pinkLine,  fill=pinkBG,  text=pinkTx},
  lbl/.style={font=\sffamily\scriptsize, text=black!55},
]

\node[dec]  (q1) at (0,0)      {Authoritative\\source?};
\node[dec]  (q2) at (0,-2.3)   {CI width?\\[1pt]{\scriptsize\mdseries best estimate}};
\node[proc] (ex) at (0,-4.2)   {Expert\\elicitation};
\node[dec]  (q3) at (0,-5.85)  {Experts\\converge?\\[1pt]{\scriptsize\mdseries ($\le$3 UF)}};

\node[tHigh] (t1)    at (5.4, 0)     {\textbf{TIER 1}\\[1pt]High confidence};
\node[tMed]  (t2med) at (5.4,-1.65)  {\textbf{TIER 2}\\[1pt]Medium confidence};
\node[tLow]  (t2low) at (5.4,-2.95)  {\textbf{TIER 2}\\[1pt]Low confidence};
\node[tLow]  (t3)    at (5.4,-5.45)  {\textbf{TIER 3}\\[1pt]Low confidence};
\node[tAb]   (t4)    at (5.4,-6.55)  {\textbf{TIER 4}\\[1pt]Abstain};

\draw[arr] (q1.east) -- (t1.west) node[lbl, midway, above] {yes};
\draw[arr] (q1.south) -- (q2.north) node[lbl, midway, right=1pt] {no};

\coordinate (f2) at ($(q2.east)+(0.55,0)$);
\draw[fork] (q2.east) -- (f2);
\draw[arr]  (f2) |- (t2med.west) node[lbl, pos=0.62, above] {$\le$2 UF};
\draw[arr]  (f2) |- (t2low.west) node[lbl, pos=0.62, below] {2--3 UF};
\draw[arr]  (q2.south) -- (ex.north)
   node[lbl, midway, right=2pt, align=left] {$>3$ UF\\(or none)};

\draw[arr] (ex.south) -- (q3.north);

\coordinate (f3) at ($(q3.east)+(0.55,0)$);
\draw[fork] (q3.east) -- (f3);
\draw[arr]  (f3) |- (t3.west) node[lbl, pos=0.62, above] {converge};
\draw[arr]  (f3) |- (t4.west) node[lbl, pos=0.62, below] {diverge};

\draw[darr] (t2low.south west) .. controls (3.0,-3.9) and (2.0,-4.0) .. (ex.east);
\node[lbl, font=\sffamily\scriptsize\itshape] at (3.15,-3.62) {elicitation optional};

\begin{scope}[shift={(-1.55,-7.5)}]
  \node[lbl, font=\sffamily\scriptsize\itshape, anchor=west] at (0,-0.45) {UF = uncertainty factor};
\end{scope}

\end{tikzpicture}%
    }
    \caption{\textbf{Determining the confidence tier.} The confidence of the classification is determined by the most uncertain contributing factor ($H$ or $E$) and tiered accordingly.}
    \label{fig:harm_estimation}
\end{figure}

\paragraph{Estimating Harm, H.} \label{sec:harm} 
Authoritative single sources rarely exist for AI harms; if they do we use them and consider them to be Tier 1 evidence (Table 2). At Tier 2, news-sourced public incident databases serve as the default for estimating lower bounds on \textit{total harm} associated with [R]. In this paper we draw on two public AI incident databases, the AI Incident Database (AIID) \cite{aiid_website} and the OECD AI Incidents and Hazards Monitor (AIM) \cite{oecd_aim_website}, which share the same basic structure (aggregating news reports on discrete incidents) and broad scope, but differ in editorial approach and sourcing methodology (see Appendix A for more details). We developed an LLM-powered script that takes an MQ and a database of incident reports as inputs, filters for full and partial matches and outputs the respective harm counts and incidents (see Appendix~\ref{app:pipeline}). Some reported entries in the AIID contain hundreds of instances of the specified harm, while others only a single one. Our pipeline handles this by attributing an estimated harm count range to each entry. These ranges are aggregated across fully matching incidents resulting in the final harm count for the respective time period. 

A partial match is an incident that matches some but not all SORT components. For example, an MQ specifying [S] as ``teenagers'' and [R] as ``LLM-assisted completed suicides'' would yield few full matches but many partial matches: adult cases that share the harm mechanism but fall short on [S], and cases of AI-assisted suicide attempts or planning falling short on [R]. A high ratio of partial to full matches suggests the MQ may be overspecified and can be revised accordingly.

Likewise for Tier 2, we construct a point estimate on the \textit{total harm} by using relevant proxy measures, while for Tier 3 we request an approximate plausible range from a domain expert. 

\paragraph{Estimating Exposure, E}\label{} We define exposure as \textit{the opportunity for harm to occur}. Exposure helps distinguish between changes in harm due to a system becoming more dangerous or being more widely deployed. Deployment and use data does not exist or is not publicly available for most AI harm types: we rarely know how many people interact with a particular system, how many decisions are automated, or how many conversations take place. As such, exposure estimation typically relies on Tier 2 methods that combine multiple partial sources. For example, combining population age data from a census with scam exposure rates from surveys would permit a point estimate of the population exposed to AI voice-clone fraud.

\FloatBarrier
\subsection{Classification}\label{sec:classification}

To classify an MQ, we first derive the harm-per-exposure trend $\hat{H}$ by comparing the rates of change of the harm~$H$ trend and exposure~$E$ trend from our estimation procedures. If~$H$ grows faster than~$E$, $\hat{H}$ is \emph{increasing}, and vice versa. MQs for which~$E$ or~$H$ cannot be determined or for which the trends are not reliably distinguishable are \emph{unclassifiable}.

The simplest non-trivial classification is a 2$\times$2-grid taking raw exposure trend~$E$ and harm-per-exposure trend $\hat{H}$ as inputs (see Figure~\ref{fig:classification}). This highlights rather than masks our epistemic limitations. The result is a mapping to four trajectories with distinct governance implications as follows:


\regime{Escalating}{\uparrow}{\uparrow\to} Both the population at risk and the harm per unit exposure are growing. This demands an urgent response: expanded monitoring, active investigation into causal drivers, and possibly regulatory intervention.

\regime{Mitigating}{\downarrow\to}{\uparrow}  More people are exposed, but harm per unit exposure is decreasing, suggesting existing safeguards are working. Continued monitoring is warranted: a failure of current controls could shift the trajectory to escalating.

\regime{Concentrating}{\uparrow\to}{\downarrow} Fewer people are exposed, but those who are face worse outcomes. This calls for targeted protective measures and investigation into why harm is intensifying.

\regime{Receding}{\downarrow}{\downarrow\to} Neither dimension is worsening. Additional intervention may not be required. Where specific measures preceded this trajectory, maintaining or extending them to related domains may be worthwhile.

\begin{figure}[h]
    \centering
    \resizebox{0.8\columnwidth}{!}{%
        \begin{tikzpicture}
\definecolor{mustard}{HTML}{EF9F27}
\definecolor{mustarddk}{HTML}{412402}
\definecolor{mustardsub}{HTML}{854F0B}
\definecolor{mustardline}{HTML}{C4A060}
\fill[mustard, opacity=0.09] (0,0) rectangle (2.8,2.0);
\fill[mustard, opacity=0.22] (2.8,0) rectangle (5.6,2.0);
\fill[mustard, opacity=0.03] (0,-2.0) rectangle (2.8,0);
\fill[mustard, opacity=0.09] (2.8,-2.0) rectangle (5.6,0);
\draw[mustardline, line width=0.4pt] (0,-2.0) rectangle (5.6,2.0);
\draw[mustardline, line width=0.3pt] (2.8,-2.0) -- (2.8,2.0);
\draw[mustardline, line width=0.3pt] (0,0) -- (5.6,0);
\node[font=\small\bfseries, text=mustarddk] at (1.4, 1.55) {Mitigating};
\node[font=\footnotesize\itshape, text=mustardsub] at (1.4, 1.05) {Monitor closely};
\node[font=\small, text=mustardsub] at (1.4, 0.35) {$\hat{H}$\,$\downarrow\rightarrow$\; E\,$\uparrow\!$};
\node[font=\small\bfseries, text=mustarddk] at (4.2, 1.55) {Escalating};
\node[font=\footnotesize\itshape, text=mustardsub] at (4.2, 1.05) {Urgent attention};
\node[font=\small, text=mustardsub] at (4.2, 0.35) {$\hat{H}$\,$\uparrow$\; E\,$\uparrow\rightarrow$};
\node[font=\small\bfseries, text=mustarddk] at (1.4, -0.45) {Receding};
\node[font=\footnotesize\itshape, text=mustardsub] at (1.4, -0.95) {Continue strategy};
\node[font=\small, text=mustardsub] at (1.4, -1.65) {$\hat{H}$\,$\downarrow$\; E\,$\!\downarrow\rightarrow$};
\node[font=\small\bfseries, text=mustarddk] at (4.2, -0.45) {Concentrating};
\node[font=\footnotesize\itshape, text=mustardsub] at (4.2, -0.95) {Targeted measures};
\node[font=\small, text=mustardsub] at (4.2, -1.65) {$\hat{H}$\,$\uparrow\rightarrow$\; E\,$\downarrow$};
\end{tikzpicture}%
    }
    \caption{\textbf{Trajectory classification.} Each monitoring question is placed in one of four categories based on the directional trends of harm-per-exposure and exposure. Monitoring questions for which either trend could not be determined with enough certainty during estimation (\S\ref{sec:estimate}) are labeled \emph{unclassifiable} and do not enter the grid. Darker fill shades correspond to increasing governance urgency.}
\label{fig:classification}
\end{figure}

\FloatBarrier
\subsection{Uncertainty quantification}\label{sec:uc}
Figure~\ref{fig:classification} reads as a deterministic grid discarding the uncertainty in the estimates behind them; in case of weak trends, even small perturbations can result in a MQ moving between cells. For each of the four estimated quantities $(\bar H_1, \bar H_2, \bar E_1, \bar E_2)$ the analyst can provide a quantification of their (or others') uncertainty. Because each quantity is positive and may range over orders of magnitude, we treat it as log-normal with median $\bar{X}$ and a central 95\% interval of $[\hat{X}/u,\, u\bar{X}]$. For the uncertainty factor $u$ holds $u\geq1$. Note that we do not provide a formal error model for determining $u$ (albeit informed by the evidence tiers of \S2.1) but only a mechanism to propagate it. As a rule of thumb, the order of magnitude of the signal has to be greater or equal to the confidence interval spanned by $u$ for a meaningful classification. Where an incident database supplies a lower bound $L_t$ on harm, the harm distribution is truncated below it (Appendix E). The four quantities are drawn independently via Monte Carlo sampling; computing both trends from the same draws preserves the correlation between exposure and harm-per-exposure, yielding a distribution over the four trajectories and Unclassifiable in place of a single cell. A trend is \emph{flat} when it lies inside an indifference band, $|\Delta E|\le\epsilon_E$ or $|\Delta \hat{H}|\le\epsilon_{\hat H}$; all results in this paper use $\epsilon_E=\epsilon_{\hat H}=0.05$. 

A trend counts as flat when the log-difference between its two estimates falls inside an indifference band $\epsilon$ of a few percent; any draw with either trend flat contributes to Unclassifiable, so epistemic indeterminacy appears as probability mass rather than as a separate rule. Setting the bands to zero and every $u = 1$ recovers the deterministic grid of Figure~\ref{fig:classification} exactly (Appendix~E).

\FloatBarrier

\section{Application and Results}\label{sec:results}

\subsection{Case Study: Conversational AI systems and self-harm.}

\noindent\textbf{Monitoring question:} [S] Among conversations between US users and conversational AI systems [O] in which users seek support regarding suicidal ideation or self-harm, [R] in how many conversations do AI systems encourage, or fail to discourage, suicidal ideation or self-harm [T] per calendar year?

\textbf{Time period:} T1: 2024 (1 January--31 December 2024); T2: 2025 (1 January--31 December 2025).

\noindent\textbf{Context:} Several high-profile news stories from 2024--2026 reported the suicides of people---often teenagers---who had confided in conversational AI systems (colloquially, ``AI chatbots''), often over the course of thousands of messages \cite{hyler2026trial}. Subsequent wrongful death lawsuits allege that the chatbots in question lacked adequate safeguards and contributed to mental health problems \cite{duffy2026character}.

``Conversational AI systems'' encompasses a broad range of systems with different functions, features, and behaviors, including: general-purpose systems (such as OpenAI's ChatGPT, Anthropic's Claude and Google's Gemini, which is integrated into Google search); persona-based AI companions (such as Character AI and Replika); and task-oriented voice assistants (such as Siri and Alexa). Although several AI chatbots have been designed by companies to improve well-being based on psychological research and expertise, as of May 2026, none has been approved by the US Food and Drug Administration (FDA) to diagnose, treat, or cure a mental health disorder \cite{abrams2025chatbots}, and the relatively cautious adoption of AI in healthcare organizations has led consumers to use general purpose AI systems as health tools by default \cite{rockhealth2026tortoise}.

\textbf{Harm estimation, H:} the analyst can choose where to begin. Starting with the lower-bound harm estimates using incident databases can offer valuable context and concrete detail, although it is worth checking first for the existence Tier 1 sources for the point estimate, which may eliminate the need to attempt lower-bound estimates.

\textbf{Lower bound:} LLM analysis of the AIID found two full matches (assessed harm count: two) in 2024 and 12 full matches (assessed harm count: 10,014--110,025) in 2025. By inspection, the 2024 matches are both individual cases of suicide, and the 2025 matches include: six cases of individual suicide, self-harm, or suicidal ideation, three reports of safety tests of conversational AI systems that had elicited responses meeting the criteria for [R], and three composite narratives. The three composite narratives are based on: a warning from the American Psychological Association (APA) regarding AI chatbots on Character.AI; an OpenAI statement regarding the number of users that show signs of suicidal ideation; and an assessment of instances of chatbot personas across several platforms deemed to  be designed to promote various harms, including suicide and self-harm. The latter two account for the four-to-five orders of magnitude increase in assessed harm count between T1 and T2.

The OECD AIM cannot be downloaded in bulk. However, its web browser offers preset filters and the option to download up to 100 results. By varying the time period, it is possible to download and concatenate multiple hundreds of results. LLM analysis of a downloaded, concatenated copy of the relevant time periods of the OECD AIM (pre-filtered for US-based incidents, involving AI ``interaction support/chatbots'' or ``content generation," resulting in death, physical injury or psychological harm), found eight full matches in 2024 and 77 full matches in 2025. By observation, most matching incidents either involve reports of actual physical self-harm or the resulting lawsuits, or are composite narratives regarding the propensity for conversational AI systems to produce undesired responses on the topics of self-harm and suicidal ideation. Removing composite narratives and de-duplicating gives one individual case of suicide in 2024 (harm count: one) and three individual cases of suicide, murder-suicide or self-harm in 2025 (harm count: four). 

Note that the results of the AIID and the OECD AIM are similar but not identical. The OECD AIM contains many more duplicate entries and composite narratives due to its sourcing methodology. The LLM assessor (used only on the AIID) can identify relevant incidents, but human inspection is helpful for validation and increased accuracy, particularly when the harm type is complex and/or has a low inclusion probability in incident databases.

\textbf{Point estimates:} We can attempt point estimates for harm, working through the methodical tiers as per the available evidence.

Tier 1: we could find no publicly available source that provided complete data on the number of conversations matching [O] and/or [R] during [T].

Tier 2: we found multiple, partially relevant (``proxy'') sources that match some of [O] and/or [R] during, or in close proximity to, [T],  from which we can extrapolate, interpolate, and otherwise combine to construct a point estimate.

Several sources provide useful context: a December 2025 survey found that 32\% of US adults had used AI chatbots for (unspecified) health information (twice as many as in 2024). Of these, around 72\% had used ChatGPT,  47\% had used Gemini and only 15\% had used a product from a healthcare provider \cite{rockhealth2026tortoise}. An October 2025 survey found that 35\% of US adults (aged 18-49) used AI tools at least once per week for mental health support \cite{ueda2026helpseeking}, and an April--May 2025 survey found that 52\% of US teens regularly used AI companions (systems designed to have conversations that feel personal and meaningful), and 12\% specifically for emotional support \cite{robb2025talktrust}

Helpfully, OpenAI published specifically-relevant figures for their products. In October 2025, they published an update to their work on preventing suicide and self-harm, in which they reported that ``around 0.15\% of users active in a given week have conversations that include explicit indicators of potential suicidal planning or intent''. The update also reports percentages of ``desired'' and ``undesired'' responses given by different models (GPT-4o, GPT-5 at initial release in August 2025, and GPT-5 as updated in October 2025) on ``challenging self harm and suicide conversations'' \cite{openai2025strengthening}. (Complications in interpreting these figures include: graphical or relative, rather than absolute, reporting of percentages; absence of pre-GPT-4o data, the differing propensity for models to degrade over long conversations---later models improve significantly---and the potential for non-default model selection by users.)

By making simplifying assumptions, we can estimate the ratio of desired-to-undesired responses given by OpenAI's ChatGPT default model on conversations matching [O]: $\sim$40:60\% January 2024--July of 2025; $\sim$80:20\% August--September 2025; $\sim$92:8\% October--December 2025.

Various sources report: the number of ChatGPT ``weekly active users'' by month (140m in January 2024, 300m in January 2025, approximately 850m in December 2025) \cite{openai2025chatgpt, backlinko2026chatgpt}; the proportions of web traffic to the main generative AI platforms by month (OpenAI's share is between $\sim$75\% and $\sim$60\%) \cite{similarweb2026tweet}; and the proportion of ChatGPT users based in the US ($\sim$18\%) \cite{explodingtopics2026chatgpt}. By linearly interpolating and extrapolating to fill in missing data, we can estimate total US-based weekly active users across all conversational AI systems each month (from $\sim$34 million in January 2024 to $\sim$243 million by December 2025).

Assuming user behavior can be generalized across the main generative AI platforms, of the total weekly active users based in the US, around 0.15\% have conversations matching [O] each week, which can be scaled to give an approximate number of monthly conversations matching [O]. (Although one user may have multiple conversations matching [O], and conversations extending beyond one week may be counted multiple times.)

Assuming that OpenAI's estimated ratios of desired-to-undesired responses can be similarly generalized, we can combine with estimates of monthly conversations matching [O] to give an estimate of the number of conversations each month that match both [O] and [R]. Summing each month for 2024 and 2025 gives approximately 2.4 million and 4 million, respectively.

\noindent\textbf{Trend:} \emph{Increasing} [$H\;\uparrow$ ]. The point estimate for T2 (4 million) is greater than the point estimate for T1 (2.4 million) by a factor of $\sim 1.7$.

\noindent\textbf{Confidence:} Medium: derived from reasonable publicly available proxy sources (Tier 2). The choice of harm type (individual conversational AI system responses) has a low inclusion probability in incident databases, except for circumstances in which conversations have led to actual physical harm. This is reflected in the prevalence of composite narratives. The lower-bound estimates, although individually unrepresentative, are directionally consistent with the point harm estimates. OpenAI explains that measuring low prevalence events is difficult as small differences in measurement can have a significant impact on the numbers reported. Nevertheless, assuming that measurement---however calibrated---remains consistent, it is clear that the overall increase in use of conversational AI systems outpaces the decrease in undesired responses due to improved safeguards.

\textbf{Exposure estimation, $E$:} Tier 1: no directly relevant and comprehensive data. Tier 2: we have already estimated the approximate number of conversations matching [O] each month as part of our point estimates for harm. Summing for 2024 and 2025 gives approximately 4 million and 12 million, respectively.

\textbf{Trend:} \emph{Increasing} [$E \uparrow$]. Exposure estimates are increasing by a factor of $\sim 3$.

\textbf{Confidence:} Medium: derived from reasonable publicly available proxy sources (Tier 2). One limitation of the simplifying assumptions that enable this analysis is aggregation bias: ChatGPT data is used as a proxy for all conversational AI platforms, and conversations are treated as equivalent regardless of user age, both of which may mask heterogeneity in the underlying patterns. As total chatbot use increased sharply throughout 2024 and 2025, OpenAI's share of users declined from around 80\% to around 60\%, such that any error in generalizing OpenAI's figures to other platforms has a proportionally greater effect on the aggregate estimate. (As of May 2026, Gemini, the second largest by share, had not released data on the proportion of their users having conversations matching [O]). The majority of high-profile cases of suicide in which AI chatbots are alleged to have played a role involve teenagers. Character platforms have substantially lower user numbers (Character AI, the largest, had 28 and 45 million monthly users in 2024 and 2025 respectively \cite{businessofapps2026characterai}), but are popular with teenagers and designed to be particularly engaging by agreeing with users and validating, rather than challenging, their thinking \cite{duane2025teenagers}. Separate and complementary analyzes might focus only on teenagers or on character platforms.

\begin{classbox}
\noindent\textbf{Classification: \textit{Mitigating}.} Harm-per-exposure, $\hat{H}\!\downarrow$ is \emph{decreasing} by a factor of around $3 / 1.7 = 0.55$, and exposure, $E \uparrow$ is \emph{increasing} by a factor of around 3. (Per-unit-exposure is less harmful, but more people are exposed; absolute harm is increasing). 

\noindent\textbf{Confidence: Medium.} (The lowest confidence from the contributing estimates.) The AI incident trajectory classifier gives the following weights, based on roughly estimated uncertainty factors of two for both harm estimates, and 1.5 for both exposure estimates: Mitigating (58.6\%), Unclassifiable (31.6\%), Escalating (9.8\%).
\end{classbox}

\subsection{LLM assessor accuracy}

To check the performance of our LLM-powered incident database assessor, we compared its results for two MQs, assessing both subject [S] and risk event [R], with those of a panel of human assessors. On the composite full-match decision (i.e. for incidents assessed to match both [S] and [R]) the LLM results were inside the human-human cloud for both MQs. However, a test based on so few points has negligible statistical significance and is more of a validation sanity check than a powered test of equivalence.

\subsection{Practitioner Demand and Applicability}

Between January and April 2026, we conducted semi-structured interviews with 42 AI risk and safety professionals spanning industry/commerce, government/civil society, and research/academia, to inform the design and assess the applicability of our methods. We found cross-sector enthusiasm for identifying clear, measurable incident types and estimating their changing impact over time (insight absent from other news and information sources). Noted applications included government research, evidence-based advocacy, governance investment justification, and updates to the EU AI Act's Code of Practice. The diversity of disciplinary approaches to AI risk and harm informed our design of the \textsc{sort} monitoring question, enforcing analytical consistency while giving full control over the incident types defined.

To quantify these observations, we administered a Likert-scale questionnaire to 44 respondents (29 of whom also participated in the consultations). 90\% rated our aims as "needed" or "greatly needed". 80\% would apply our methods directly, and 95\% would use the results of others if methodologically sound, suggesting the existence of a latent analytical community. Detailed findings by stakeholder type and survey results are documented in Appendix~C.

\section{Discussion}\label{sec:discussion}

This paper proposes a simple methodology for defining and comparing AI-related harms between two (typically subsequent) time periods (T1 and T2). A monitoring question (MQ), based on the SORT (Subject, Opportunity, Risk event, Timeframe) framework, defines the scope of study of both harm and exposure. Point estimates of harm and exposure are constructed systematically from publicly-available data sources, supported where necessary by expert judgment. Incident databases provide a lower-bound estimate of harm as well as rich contextual detail to help define an appropriate MQ and interpret resulting classifications. Exposure contextualizes the harm estimate, helping to distinguish “more harmful” from “more deployed/used” AI systems. Data sources and methods for their analysis are grouped into “tiers” by type according to the approximate level of confidence in the accuracy of resultant estimates (subject to user judgment and adjustment). A statistical model transforms the harm and exposure point estimates (as modified by the lower-bound harm estimate and user-judged uncertainty factors), into an assessment of the probability of the change between T1 and T2 falling into one of four possible classifications (\emph{Escalating, Mitigating, Concentrating, Receding}) or being deemed \emph{Unclassifiable}. Each of the four classifications describe a “trajectory” between two time periods by expressing harm and exposure in T2 relative to T1.

The case study on conversational AI and self-harm illustrates the importance of accounting for exposure. Incidents of harm are estimated to increase between T1 and T2, such that a “naive” interpretation might conclude that conversational AI systems are becoming riskier. However, deployment and use is estimated to increase significantly faster, suggesting that, although absolute harm is increasing, conversational AI systems are becoming safer per use. 

The classification tells us nothing about the absolute scale of harm and should therefore always be interpreted alongside the underlying estimates. A \emph{Mitigating} classification can coexist with large and growing absolute harm if exposure rises faster than harm-per-exposure falls. A \emph{Receding} classification may still represent a substantial burden. A \emph{Concentrating} classification at high absolute levels may warrant more urgent attention than an \emph{Escalating} classification at low levels. Full analyses must consider net harm i.e. the AI-related harm net of any harm avoided by displacing other causes of harm.

Changes in classification between pairs of time periods (e.g. from T1 to T2 then from T2 to T3) can support governance learning. A shift from \emph{Escalating} to \emph{Mitigating} may indicate a successful intervention, whereas a shift from \emph{Receding} to \emph{Concentrating} may signal deepening harm within a subpopulation, requiring an adjustment to an existing response. In principle, sustained application of the classification process over time should allow worsening trajectories to be caught before harm scales. However, at present the availability of data is a significant constraint on the methodology, such that it would be challenging to fit data neatly to the span of a policy intervention, let alone anticipate a governance response.

Where confidence in data sources (and therefore in estimates) is high, the relative dominance of harm or exposure in a classification can inform the intervention type: exposure-dominant trajectories might respond best to measures designed to control deployment or access, whereas harm-dominant trajectories require focus on the harm mechanism itself.

Where confidence in data sources is low, classifications have an advantage in robustness over individual harm or exposure estimates, as stable sources of bias and uncertainty cancel out when compared across time periods (assuming a fixed measurement protocol).Even when data sources are inadequate and expert support is insufficient to reach any meaningful conclusion, setting up and attempting to answer an MQ can reveal gaps in data for known harms, such that \emph{Unclassifiable} can be a useful in guiding policy action (i.e. supporting further monitoring efforts).



At current incident reporting levels, the effect of the lower bound harm estimate on any classification is typically limited. In principle, it  “clips” the lower portion of the uncertainty bound around the point estimate of harm, which is used to generate the probabilistic assessment of each possible classification for a given MQ. As such, the relevance of the lower bound harm estimate increases as its value approaches that of the point estimate. To put this another way: the better the incident data, the more our methodology uses and relies on it.

Although current and forthcoming regulatory regimes are unlikely to produce a significant increase in the volume of publicly-available incident data, demonstrating the potential for governance insights through simple analysis may encourage more systematic and comprehensive reporting. Additionally, the framework is applicable for a wide range of incident types (as demonstrated with more MQ), which can proof useful also risk management within organizations or government departments. 

That said, certain types or levels of harm exceed the capacity of individual incident reporting. For example, when a harm reaches a prevalence easily detectable at population-level (e.g., non-consensual sexual deepfakes), or is too diffuse, private or personal to be reported (e.g., our case study on conversational AI systems and self-harm), prevalence surveys and data from system and platform providers are needed instead.

Several limitations remain. The binding one is data availability: for many harm types no source supports a classification between two periods, and where data exist they are often of unknown quality and coarse granularity, forcing the analysis onto the collection period rather than the period of interest. The classifier is also blind within a period: it reads a spike-and-reversal as monotonic change.

Intentional harms in particular challenge our methods, and our appendix shows why. The framework assumes risk accrues passively through recorded deployment and use; attackers declare nothing, and failed attacks go unreported. The consequences are visible across three worked cases: cyberattacks (Appendix D2) and deepfake scams (Appendix D3) resolve as unclassifiable, while voice-clone fraud (Appendix D4) is classifiable only against an exposure proxy defined at the level of $[S]$ rather than the population actually targeted, a rather weak denominator. The analyst's choice is between this easy-but-flat $[S]$ estimate and the harder attacked-population estimate; only the latter speaks to the harm of interest, and even a crude version of it would inform governance and expose data gaps.

In conclusion: the monitoring of AI harms relies currently on raw incident counts from news-sourced databases without standardized definitions, methodologies, or adjustment for exposure or deployment scale \cite{bengio2026iaisr, saeri2026prioritization, rao2026challenges}. This paper proposes a simple methodology to fill these gaps. Applying it will not identify emergent harm mechanisms, but will reveal weaknesses in current reporting infrastructure, clarify where data collection needs to improve and, hopefully, help demonstrate the prospective value, and motivate the introduction, of systematic incident reporting regimes.

\section*{Impact Statement}

The framework replaces the current path from raw counts through naive trend claims to reactive governance with decomposed trends, trajectory classification, and prioritisation grounded in exposure-adjusted harm. Current reporting regimes are narrowly scoped and unlikely to produce sufficient data on their own; the framework is designed to work with the news-sourced databases that remain the primary record of AI harm, and to demonstrate the governance value that could motivate broader reporting infrastructure over time. It makes visible whether rising counts reflect genuine risk escalation due to more harmful or more widely deployed AI systems, or simply increased media interest, and where reporting infrastructure is insufficient to support trend claims

\section*{Generative AI Use}
Large language models were used to assist with editing, formatting, and manuscript preparation. Additionally, an LLM is an integral component of the incident-database assessor tool central to our methodology, and LLMs assisted with the code implementing it; these uses are described in the main text and Appendix~\ref{app:pipeline}. All conceptualization, research design, analysis, and interpretation are the authors' own. All AI-assisted text and outputs were reviewed and verified by the authors, who take full responsibility for the content of the paper.

\bibliography{references}
\bibliographystyle{icml2026}

\newpage
\appendix
\onecolumn
\section{Appendix}
The pathway to inclusion in a public incident database can be thought of as a funnel with sequential filters through which an incident must pass:
\begin{itemize}
    \item Detection: the harm occurs and is detectable
    \item Attribution: the harm is recognised as an AI failure
    \item Recording: a record of the harm is created and survives
    \item Disclosure and reporting: the record reaches someone who reports it
    \item Capture: the database picks up the report
    \item Within scope: the report fits the database's definitional scope
\end{itemize}
\subsection{Systematic biases in incident reporting}
A range of biases (including visibility, detection, disclosure incentive, media salience, geography/language and victim voice) operate across the stages and compound multiplicatively. This makes certain types of incidents more likely to be included than others:
\begin{description}
    \item[High inclusion probability:] Acute, dramatic, novel, unambiguously AI-related, consumer-facing harm to well-resourced, individual (rather than systemic), English-language, vocal victim(s), in regulated sectors and fashionable domains. E.g.\ a US-based LLM chatbot produces a shocking output, goes viral on social media, and the company issues a public statement; an autonomous vehicle is involved in a fatal collision investigated by a regulator.
    \item[Moderate inclusion probability:] Incidents that are detectable and recorded but face friction at the attribution or disclosure filter. E.g.\ algorithmic bias cases pursued through litigation, incidents surfaced by academic surveys rather than the media, terrorism-related incidents which are suppressed in the interests of national security.
    \item[Low inclusion probability:] Slow, subtle, chronic, diffuse, systemic harm, harder to attribute, affecting marginalised or geographically remote populations, in unregulated domains, with non-disclosure incentives and no or limited institutional record. E.g.\ a credit-scoring model in a non-English-speaking country penalises residents of certain postcodes; harm is real but diffuse, no individual claimant has the resources to pursue it, no journalist reports it, and the system is eventually replaced without the pattern ever being formally named.
\end{description}

\subsection{Incident databases in practice: the AIID and OECD AIM}
The AI Incident Database (AIID) \cite{aiid_website} and the OECD AI Incidents and Hazards Monitor (AIM) \cite{oecd_aim_website} are among the most well known and broadly populated public AI incident databases. They have similar basic structures (one overarching incident title aggregates content---predominantly news articles---from multiple outlets on the same incident) but diverge somewhat in their editorial approach and sourcing methodology, an understanding of which is helpful in contextualising their use in harm classification.

The \textit{AIID Editor's Guide} \cite{aiid_editors_guide} defines an AI incident as: ``an alleged harm or near harm event to people, property, or the environment where an AI system is implicated.''

The \textit{AIM Overview and Methodology} \cite{oecd2024aim} defines an AI incident as ``an event, circumstance or series of events where the development, use or malfunction of one or more AI systems directly or indirectly leads to [a specific set of] harms.'' The methodology also defines an AI hazard separately as ``an event, circumstance or series of events where the development, use or malfunction of one or more AI systems could plausibly lead to an AI incident.'' (The database can be filtered to show either or both of these.)

At the time of writing, the AIID recorded 1,460 incidents, and the OECD AIM recorded 9,218 incidents and 5,312 hazards (14,530 incidents and hazards).

The AIM sources its data in the form of clusters of articles reporting on the same AI-related event (not pre-filtered for harm or hazard) from a news intelligence platform. It uses LLMs to classify and filter events as incidents, hazards, or unrelated content. It uses a rolling four-day processing window, such that related articles appearing more than four days apart risk misclassification as separate events. In practice, the filters also capture reports that do not meet the definition of a discrete incident, including composite narratives, pattern-explanation stories, and accounts of potential vulnerabilities.

The AIID sources its data via submissions from a community of contributors. At the time of writing, the AIID has received a total of 5,866 submissions: 4,932 from 123 named contributors, and 934 from an unknown number of anonymous contributors. Submissions primarily comprise news articles, along with academic articles, legal filings, and social media posts. Correspondence between the AIID and contributors indicates that the majority source submissions passively (i.e.,\ reading an article describing an incident, then submitting it), with some also using active methods including keyword-based alerts and searches, and LLM and Google Translate-supported coverage of non-English sources. Upon submission, the editorial process is human-mediated, with an ethos of comprehensive collection and precise filtering for discrete incidents, or in some cases clusters of multiple incidents aggregated at the title and description level.

The two databases likely draw on substantially overlapping, though not identical, sources, and are shaped by common patterns of source visibility and by constraints inherent to news-based reporting. Both databases are valuable resources in their own right, and together offer complementary coverage of how and to whom AI systems cause harm.

Researchers using either database to count incidents should be attentive to how each database is constructed, in particular:
\begin{itemize}
    \item The two databases define incidents slightly differently, though the SORT Framework's precise specification of the incident type of interest should help navigate this.
    \item In practice, news articles about the same incident are often published days, months, and even years after the event, as initial reports give way to analytical pieces and trend stories, meaning the AIM's four-day processing window likely has a real impact on its incident counts, with single incidents fragmented across multiple entries.
    \item The AIM's tendency to capture composite narratives alongside discrete incidents, and the AIID's occasional use of cluster-level incident titles---when coverage treats closely related events as part of a single unfolding story---reflect the difficulty of translating complex reporting trajectories into clean incident counts.
\end{itemize}
That said, given the current systemic underreporting of AI-related harms, even a database that overcounts some incidents is likely to represent a reasonable lower bound for most forms of harm.

\FloatBarrier
\section{Appendix}

\section*{Consultation Methodology}

Between January and April 2026, we conducted semi-structured consultations (31 video interviews, 11 written responses) with 42 AI risk and safety professionals across 9 sectors (see Table~\ref{tab:respondents}). Participants were given an overview of our research aims and asked: whether and how they currently monitor AI risks and incidents; what they found most difficult about staying informed; how useful they would find our proposed classification methods; and what they would need for the outputs to be applicable to their work. These consultations informed the design of our classification methods and helped assess their potential utility and applicability across a range of professional contexts and disciplinary perspectives on AI risk and harm.

\begin{table}[ht]
\centering
\caption{Consultation and survey participant role, sector and count}
\label{tab:respondents}
\begin{tabular}{p{1.2cm} p{4.5cm} p{7cm} p{1.5cm}}
\hline
\textbf{Survey count} & \textbf{Sector} & \textbf{Interviewee/respondent role}& \textbf{Interview count} \\
\hline
9 & Academia & AI policy researcher; AI safety doctoral student; AI safety professor; cybersecurity postgraduate; security professor; undergraduate AI research student & 6 \\
5 & AI technical research/testing (non-governmental) & AI safety researcher (2); AI security founder & 3 \\
7 & Civil society, NGO, think tank & AI governance specialist; AI policy advisor; AI risk manager; AI safety advocate; AI safety founder; chief AI officer; research director & 7 \\
9 & Consulting/advisory/independent & AI governance consultant (3); AI governance manager; AI risk advisor; AI risk consultant; AI security consultant; data privacy consultant; organisational psychologist; risk consultant; risk/audit consultant & 10 \\
9 & Financial services (banking, insurance, payments) & AI governance auditor; AI governance lead; AI security lead; head of analytics; incident manager; risk manager; security officer & 7 \\
4 & Government policy, research or regulation & AI governance advisor; AI risk advisor (2); AI risk researcher; AI testing coordinator; regulatory researcher & 5 \\
0 & Healthcare & AI security architect; AI strategist & 2 \\
0 & Media and communications & AI safety journalist & 1 \\
1 & Technology & AI project lead & 1 \\
\hline
44 & 9 sectors & 40 interviewee/respondent roles & 42 \\
\hline
\end{tabular}
\end{table}

Following the consultation phase, we administered a structured five-point Likert-scale questionnaire to 44 respondents (29 of whom also participated in the consultations), covering current practices and experiences of staying informed about AI risk and harm, the perceived need for our project aims and willingness and ability to apply our methods. (Our project aims were described as: identifying clear, measurable incident types and estimating their frequency, severity, and change over time.)

Survey results were grouped by sector to improve statistical significance and reflect broadly similar conceptualisations of AI risk and harm:
\begin{itemize}
    \item through operational controls and compliance (``industry and commercial'', $n=19$, combining financial services, consulting and technology);
    \item through empirical research and capability testing (``research and academia'', $n=14$, combining academia and non-governmental AI technical research and testing);
    \item through policy, advocacy and regulation (``policy and civil society'', $n=11$, combining civil society organisations and government policy, research and regulation).
\end{itemize}

%
\definecolor{sapph50}{HTML}{E6F1FB}
\definecolor{sapph100}{HTML}{B5D4F4}
\definecolor{sapph200}{HTML}{85B7EB}
\definecolor{sapph400}{HTML}{378ADD}
\definecolor{sapph600}{HTML}{185FA5}
\definecolor{sapph900}{HTML}{042C53}
\newcommand{\hA}[1]{\cellcolor{sapph50}#1}
\newcommand{\hB}[1]{\cellcolor{sapph100}#1}
\newcommand{\hC}[1]{\cellcolor{sapph200}#1}
\newcommand{\hD}[1]{\cellcolor{sapph400}\textcolor{sapph50}{#1}}
\newcommand{\hE}[1]{\cellcolor{sapph600}\textcolor{sapph50}{#1}}

\begin{table}[!ht]
\centering
\small
\caption{Aggregated Likert responses by sector group (n = 44).}
\label{tab:appC-sector}
\begin{tabular}{@{}lccccc@{}}
\toprule
Group & 1 & 2 & 3 & 4 & 5 \\
\midrule
\multicolumn{6}{@{}l}{\textit{How much do you need to stay up to date with developments in AI risk and harm to do your job?}} \\
\multicolumn{6}{@{}l}{\scriptsize (1 = not at all, 5 = very much)} \\
Industry/commercial (n=19) & 0 (0\%) & \hA{1 (5\%)} & \hA{2 (11\%)} & \hB{4 (21\%)} & \hE{12 (63\%)} \\
Research/academic (n=14) & 0 (0\%) & 0 (0\%) & 0 (0\%) & \hB{3 (21\%)} & \hE{11 (79\%)} \\
Policy/civil society (n=11) & 0 (0\%) & 0 (0\%) & 0 (0\%) & 0 (0\%) & \hE{11 (100\%)} \\
\addlinespace[6pt]
\midrule
\addlinespace[4pt]
\multicolumn{6}{@{}l}{\textit{How hard is it to know which AI risks/harms have the greatest impact?}} \\
\multicolumn{6}{@{}l}{\scriptsize (1 = very hard, 5 = very easy)} \\
Industry/commercial (n=19) & \hA{1 (5\%)} & \hB{5 (26\%)} & \hB{3 (16\%)} & \hC{7 (37\%)} & \hB{3 (16\%)} \\
Research/academic (n=14) & \hA{2 (14\%)} & \hA{2 (14\%)} & \hC{6 (43\%)} & \hB{4 (29\%)} & 0 (0\%) \\
Policy/civil society (n=11) & \hB{2 (18\%)} & \hA{1 (9\%)} & \hC{4 (36\%)} & \hC{4 (36\%)} & 0 (0\%) \\
\addlinespace[6pt]
\midrule
\addlinespace[4pt]
\multicolumn{6}{@{}l}{\textit{How hard is it to know which AI risks/harms are increasing in frequency or severity?}} \\
\multicolumn{6}{@{}l}{\scriptsize (1 = very hard, 5 = very easy)} \\
Industry/commercial (n=19) & \hA{1 (5\%)} & \hB{4 (21\%)} & \hB{4 (21\%)} & \hC{6 (32\%)} & \hB{4 (21\%)} \\
Research/academic (n=14) & \hA{2 (14\%)} & \hA{2 (14\%)} & \hB{3 (21\%)} & \hB{4 (29\%)} & \hB{3 (21\%)} \\
Policy/civil society (n=11) & \hA{1 (9\%)} & \hB{3 (27\%)} & \hA{1 (9\%)} & \hC{5 (45\%)} & \hA{1 (9\%)} \\
\addlinespace[6pt]
\midrule
\addlinespace[4pt]
\multicolumn{6}{@{}l}{\textit{How needed are better methods for estimating the severity/impact of different types of AI incident?}} \\
\multicolumn{6}{@{}l}{\scriptsize (1 = not needed, 5 = greatly needed)} \\
Industry/commercial (n=19) & 0 (0\%) & 0 (0\%) & \hA{1 (5\%)} & \hC{6 (32\%)} & \hE{12 (63\%)} \\
Research/academic (n=14) & 0 (0\%) & 0 (0\%) & \hA{2 (14\%)} & \hB{4 (29\%)} & \hD{8 (57\%)} \\
Policy/civil society (n=11) & 0 (0\%) & 0 (0\%) & 0 (0\%) & \hB{3 (27\%)} & \hE{8 (73\%)} \\
\addlinespace[6pt]
\midrule
\addlinespace[4pt]
\multicolumn{6}{@{}l}{\textit{How needed are better methods for estimating how different types of AI incident are changing over time?}} \\
\multicolumn{6}{@{}l}{\scriptsize (1 = not needed, 5 = greatly needed)} \\
Industry/commercial (n=19) & 0 (0\%) & 0 (0\%) & \hA{1 (5\%)} & \hB{4 (21\%)} & \hE{14 (74\%)} \\
Research/academic (n=14) & 0 (0\%) & 0 (0\%) & \hA{2 (14\%)} & \hB{4 (29\%)} & \hD{8 (57\%)} \\
Policy/civil society (n=11) & 0 (0\%) & 0 (0\%) & \hA{1 (9\%)} & \hB{3 (27\%)} & \hE{7 (64\%)} \\
\bottomrule
\end{tabular}
\end{table}
\FloatBarrier

\subsection*{Overall Results}

Enthusiasm for the project aims was broad across all sectors and role types in consultations. Survey respondents rated all aims highly, with a majority giving the maximum score ``greatly needed'', and no aim rated as ``not needed''.

Interviewees across sectors described current reliance on news alerts, personal networks and ad hoc searches, and noted that reliably establishing whether a specific incident type is increasing or decreasing in frequency or harm is difficult by these means. The two tasks survey respondents found hardest---knowing which risks are increasing and decreasing---precisely match our project aims.

Outside of the specific project aims, access to more incident data, and the ability to filter and sort it according to requirements, was the most frequently requested feature. Despite the existence of various industry and cross-sector standards and frameworks there is no universal taxonomy or common articulation of incident types. AI risk and harm are so domain-specific and sociotechnically complex that practitioners' ways of thinking about them are as varied as their professional contexts. This informed our design choice of the SORT monitoring question: it provides analysts with a consistent structure for meaningful searching while giving them full control over the incident types they define.

Despite being informed that a single monitoring question takes one to two hours to evaluate, 35 respondents (79.5\%) said they would have time and 37 (84.1\%) the confidence to implement the methods themselves, with 42 (95.5\%) saying they would also use results on a dashboard, if they could be trusted. Trust would require: transparent and reproducible methodology, clearly defined inclusion criteria, explicit acknowledgement of reporting bias and data gaps, and consistency over time.

\subsection*{Limitations and Caveats}

Interviewees for whom the methods were unlikely to be directly practical were either working upstream of incidents in technical threat modelling or AI safety testing, where decisions and priorities are driven by model capabilities and deployment timing, or in roles sufficiently structured by existing frameworks, policies and controls that incident tracking was a secondary concern.

Several interviewees raised concerns about the scope of what incident databases can capture---points generally reflected in our limitations---showing the importance of communicating clearly to prospective users to avoid misinterpretation of what incident trend data can and cannot show.

\subsection*{By Sector}

\subsubsection*{Financial Services}

Interviewees were consistently enthusiastic, the most focused on current (rather than prospective) harms, and the most acutely aware of the monitoring gap. A lack of reliable incident tracking or cross-organisational classification consistency resulted in compliance teams reliant on informally shared news links. Where available, sector-specific trend data strengthens the case for security reviews and governance investment and can inform insurance underwriting. Financial services survey respondents rated tracking change over time highest of any sector, while reporting the greatest difficulty with current monitoring.

\subsubsection*{Consultants, Auditors and Governance Advisors}

Enthusiasm centred on justifying governance investment and standardising classification. Linking trends to established frameworks---ISO 42001, NIST RMF, and IEEE standards---would make action explicit, though evaluation periods may be longer than practitioners need given current data availability.

\subsubsection*{Civil Society, Advocacy and NGOs}

Interviewees were among the most enthusiastic for support to evidence-based advocacy, but also the most capacity constrained. Quantifying harm is valuable for planning, prioritisation and communication. Demonstrating increasing harm would make advocacy substantially more persuasive to the small number of well-placed decision-makers who can meaningfully reduce AI harms. Granularity in trend data dictates the resolution at which advocacy can be targeted, so any improvement could have second order effects. An EU policy advisor observed that better tools are needed for monitoring how risks change over time, and that a centralised source of such changes would directly support updates to the EU AI Act's Code of Practice.

\subsubsection*{Government Policy, Research and Regulation}

Enthusiasm was strong. Given government interest in the whole spectrum of AI risks, even a lower bound on incident prevalence would be a widely useful tool for risk assessments and inter-departmental communications. Estimating the prevalence of specific issues is also a low cost way to scope new research.

\subsubsection*{Academic and Independent Researchers}

Researchers and independent practitioners are the most willing and able to implement the methods themselves. Enthusiasm centred on structure and pattern recognition: the potential to transform fragmented observations into coherent, evidence-based narratives. A technical AI safety researcher noted the potential for reproducible classifications to be cited in legal filings.

\subsubsection*{Healthcare}

Incident trend data would support regulatory submissions requiring consideration of the magnitude of potential harm, and supplement frameworks that provide risk data but do not rank by prevalence or trajectory.

\FloatBarrier
\section{Appendix}
\subsection{LLM Assessor Pipeline}
\label{app:pipeline}

\begin{algorithm}[H]
  \caption{LLM-based harm assessment (lower-bound estimate)}
  \label{alg:harm_pipeline}
  
  \KwIn{Monitoring question (SORT); an incident database; two time periods}
  
  \KwOut{A lower-bound harm trend; per-incident assessments}
  
  \ForEach{incident in the database}{
    Assess subject and risk-event match as \textsc{true}, \textsc{false}, or \textsc{indeterminate}
    
    Extract harm quantity 
  }
  \nl Separate full matches from partial matches (both \textsc{true} or any \textsc{indeterminate})
  
  \nl Compute harm totals for each period
  
  \uIf{fewer than 3 full matches in either period}{
    Abstain (insufficient evidence)\;
  }

  \Else{
    Report increasing or decreasing accordingly\;
  }
  \If{partial matches greatly outnumber full matches}{
    Flag: monitoring question may be overspecified\;
  }
\end{algorithm}

Algorithm~1 is implemented as three decoupled stages---loader, assessor, aggregator---driven by a single MQ config containing the SORT tuple $(S,R)$ and either explicit comparison periods or a reporting frequency $\in\{\text{monthly, quarterly, yearly}\}$.

\paragraph{Loader.} Supports the AIID snapshot (1{,}405 incidents, 6{,}787 reports) and OECD AIM exports. When a frequency is given, periods are the two most recent complete periods at that cadence, offset by a 3-month reporting buffer from the run date. By default the assessor receives only the incident's title, description, and alleged deployer/developer/harmed parties; optionally, the first linked report with non-empty text can be attached, truncated to 4000 characters.

\paragraph{Assessor.} Per incident we call the OpenRouter Chat Completions API (default \texttt{anthropic/claude-3.5-haiku}; any OpenRouter model is substitutable and the exact string is logged per run) at temperature $0.1$, JSON response mode, and a 1000-token cap. The prompt embeds the SORT fields and incident record and requires a flat JSON object with: $S\_match, R\_match \in \{\textsc{true}, \textsc{false}, \textsc{indeterminate}\}$ with short reasoning; integer harm bounds $[\text{lower},\text{upper}]$ in the harm unit with reasoning; and a suggested proxy measure. The model is instructed to prefer \textsc{indeterminate} over \textsc{false} when in doubt. Up to 3 attempts with exponential backoff and longer waits on rate-limit errors; exhausted retries yield both matches \textsc{indeterminate}. Calls can be parallelised, and a single invocation can batch a list of MQs.

\paragraph{Aggregator.} An incident is a \emph{full match} if $S\_match=R\_match=\textsc{true}$, and a \emph{partial match} if either component is \textsc{indeterminate}. Per-period bounds are unweighted sums over full matches, $H^{\text{lower}}_p=\sum_{i\in\text{Full}(p)} \text{lower}_i$ and analogously for the upper bound including partial matches.

\paragraph{Inter-assessor agreement.}
We validate pipeline output against human raters on two MQs (AV injury, AI chatbots), each scored once with \texttt{claude-3.5-haiku}. For each MQ, the sampler stratifies pipeline output into full/partial/negative and draws a seeded random sample of up to $k$ per stratum; raters label a blinded, shuffled subset using the same schema as the pipeline (\texttt{S\_match}, \texttt{R\_match}, severity, harm-quantity bounds). We report Cohen's $\kappa$ on \texttt{S\_match} (3-way), \texttt{R\_match} (3-way), and the binary full-match for all pairs of raters, treating the LLM as one rater among humans. This is a single-model, single-run evaluation: run-to-run stability of the same LLM and cross-model agreement are out of scope and left to future work.

\begin{figure}[h]
    \centering
    \includegraphics[width=\linewidth]{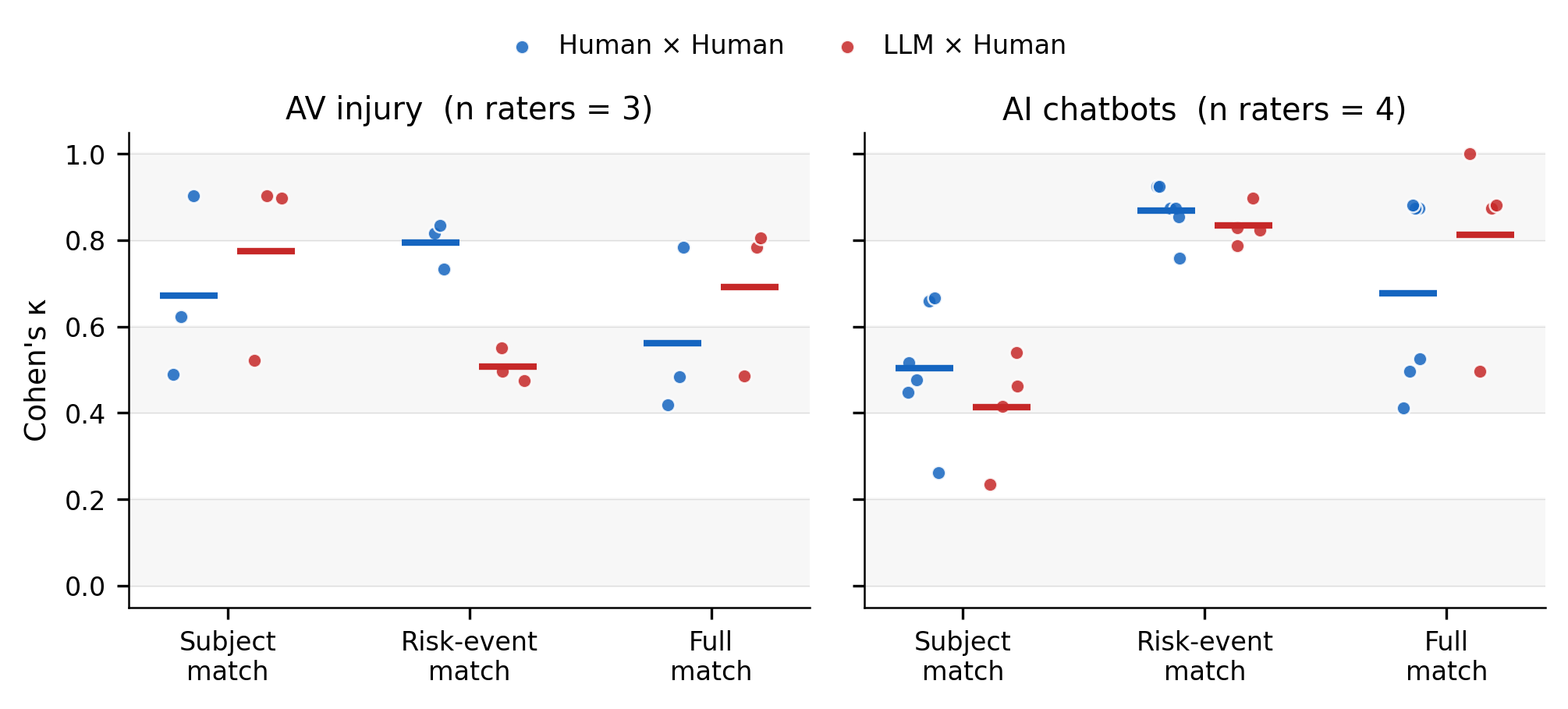}
    \caption{Inter-rater agreement (Cohen's $\kappa$) for each topic. Dots show pairwise $\kappa$ between raters on the 3-valued S\_match / R\_match labels and the derived binary full-match; horizontal bars mark group means. Human--human pairs define the achievable ceiling; LLM--human pairs within that cluster indicate model agreement at human-panel quality. Shaded bands: Landis \& Koch interpretation zones.}
    \label{fig:inter-rater}
\end{figure}

\paragraph{Results.}
Three raters scored AV injury and four scored AI chatbots. Figure~\ref{fig:inter-rater} plots pairwise $\kappa$ for every human-human pair (the ceiling implied by human disagreement) and every LLM-human pair. On the composite full-match decision the LLM sits inside the human-human cloud on both topics ($\bar{\kappa}_{\text{LH}} = 0.69$ vs.\ $\bar{\kappa}_{\text{HH}} = 0.56$ for AV injury; $0.81$ vs.\ $0.68$ for AI chatbots). The decomposed labels show topic-specific asymmetries: on AV injury the model matches or beats humans on subject identification ($0.77$ vs.\ $0.67$) but falls below the ceiling on risk-event identification ($0.51$ vs.\ $0.80$); on AI chatbots the pattern reverses---the model matches humans on risk-event ($0.83$ vs.\ $0.87$) and is slightly below on subject ($0.41$ vs.\ $0.51$).

\paragraph{Validity.}
With 3-6 human-human and 3-4 LLM--human pairs per topic, this is a validation sanity-check, not a powered test of equivalence: a nonparametric comparison of $\bar{\kappa}_{\text{HH}}$ and $\bar{\kappa}_{\text{LH}}$ from so few points has negligible statistical power. One LLM-human pair on AI chatbots yielded $\kappa = 1.0$ on full-match, a degree of agreement not seen in any other pair and consistent with possible label leakage during rating; we plot it for transparency but do not rely on it. Because we evaluated a single model on a single run, these results speak to agreement with human judgment on two MQs and not to the stability of that agreement across runs or models; a stronger validation would require at least a second independent run and a second model, plus $\geq 5$ complete raters per topic with pre-registered bootstrap CIs on $\bar{\kappa}_{\text{LH}} - \bar{\kappa}_{\text{HH}}$.

\section{Appendix}\label{App:Examples}

\subsection{Autonomous vehicles and injury or property damage.}

\noindent\textbf{Monitoring question:} [S] Among autonomous vehicles (SAE Levels 3 through 5) [O], per million miles driven on US public roads, [R] how many incidents involving injury or property damage occur [T] per calendar year?

\textbf{Time period:} T1: 2024 (1 January--31 December 2024); T2: 2025 (1 January--31 December 2025).

\textbf{Context:} The Society of Automotive Engineers (SAE) classifies systems into six levels from 0 (no driving automation) to 5 (full driving automation). From 2021, the US National Highway Traffic Safety Administration (NHTSA) requires specified manufacturers and operators to report certain crashes involving vehicles equipped with automated driving systems (ADS) or SAE Level 2 advanced driver assistance systems (ADAS). ADS-enabled entities must report crashes in which ADS was in use within 30 seconds of the crash and which resulted in specified property damage, collision with a vulnerable road user, airbag deployment, tow-away or the hospitalization or fatality of any injured party.

\textbf{Harm estimation, H:} Tier 1: the NHTSA records 537 and 1,021 ADS incidents matching [O] and [R] in 2024 and 2025 respectively (primarily property, damage rather than injury) \cite{nhtsa2023sgo}. Where the analyst has access to data at Tier 1, there is no need to attempt a lower-bound harm estimate, unless of interest for context and detail.

\noindent\textbf{Trend:} \emph{Increasing} [$H\;\uparrow$ ]. By a factor of around 1.9.

\noindent\textbf{Confidence:} High: derived from authoritative and directly relevant data (Tier 1). Although the NHTSA cites certain data limitations: initial reports may reflect incomplete or unknown information; the same crash may have multiple reports if reported by multiple entities, or if new information becomes available and a new report is filed.

\textbf{Exposure estimation, $E$:} Tier 2: since 2023, the US Autonomous Vehicle Industry Association (AVIA) has published an annual estimate of the total number of miles driven by autonomous vehicles (AVs) in the US: 44 million by around June 2023, 70 million by June 2024 and 145 million by May 2025 \cite{avia2025}. The time period varies year-to-year and does not align with the MQ time period, however, linear interpolation allows us to estimate total miles driven by AVs in 2024 ($\sim57$ million). To estimate total miles driven in 2025 we need to extrapolate beyond the available AVIA data. The AV industry is expanding rapidly, such that linear extrapolation from the AVIA figures would produce an underestimate. We can combine several sources to increase accuracy. Waymo is the largest US source of AV miles (accounting for 68\% of the NHTSA accident reports \cite{nhtsa2023sgo}, and around 49\% of the AVIA total miles driven as of May 2025) and although they do not publish regular updates, they have publicized key milestones, including 100m and 200m total autonomous miles driven in July 2025 and February 2026 respectively \cite{waymo2025hundred, waymo2026twohundred}. Linear interpolation of Waymo's own figures give $\sim170$ million total miles driven by December 2025. Scaling this figure by a simple average of the two estimates for Waymo's share of the AV market (49\% and 68\%) allows us to estimate the total all-time AV miles driven ($\sim$300 million) and hence the total AV miles driven in 2025 ($\sim187$ million).

\textbf{Trend:} \emph{Increasing} [$E \uparrow$]. By a factor of around 3.3.

\textbf{Confidence:} Medium: derived from reasonable proxy sources (Tier 2).

\begin{classbox}
\textbf{Classification: \textit{Mitigating.}} Harm-per-exposure, $\hat{H}\!\downarrow$ is \emph{decreasing} by a factor of around 0.58 (from around 9.4 incidents per million miles in 2024, to 5.4 incidents per million miles in 2025), whereas exposure, $E \uparrow$ is \emph{increasing} by a factor of around 3.3. (Per-unit-exposure is less harmful, but more people are exposed and absolute harm is increasing. However, a full analysis should consider net harm, i.e., ADS-attributable incidents net of the harm avoided by displacing non-autonomous vehicles.).

\noindent\textbf{Confidence: Medium.} (The lowest confidence from the contributing estimates.) The AI incident trajectory classifier gives the following probabilities, based on roughly estimated uncertainty factors of 1.1 for both harm estimates, and 1.5 for both exposure estimates: Mitigating (91\%), Unclassifiable (6.9\%), Escalating (2.1\%).
\end{classbox}

\subsection{AI-enabled cyberattacks on large financial services organizations.} \label{d2}

\noindent\textbf{Monitoring question:} [S] Among financial services organizations operating globally with 1{,}000+ employees, [O] that are attacked using AI-enabled cyberattacks [R] how many suffer data breaches as a result [T] per calendar year?

\noindent\textbf{Time period:} T1: 1 November 2023--31 October 2024;
T2: 1 November 2024--31 October 2025. (Offset from calendar year to accommodate data availability.)

\noindent\textbf{Context:} The financial services sector has long been heavily targeted by financially motivated cyber attackers \cite{verizon2026dbir}. Since ChatGPT's release in November 2022, the use of generative AI in cyberattacks has been anticipated by cybersecurity researchers, but until early 2024, little concrete evidence had been reported \cite{verizon2024dbir, ibm2024xforce}. By mid-to-late 2024, this had shifted: the proportion of malicious emails written by AI had doubled. OpenAI and Google Threat Intelligence Group (GTIG), who monitor their models to investigate and disrupt suspected malicious activity, reported attempts to use ChatGPT and Gemini in influence operations, phishing, malware debugging, and social engineering---primarily for productivity gains, rather than novel capabilities \cite{openai2024deceptive, openai2024influence, google2025adversarial, verizon2025dbir}. Use escalated throughout 2025: GTIG observed the first instances of LLMs embedded directly in malware execution; the emergence of an underground marketplace for ``AI attack tooling''; and the continued augmentation of the ``full attack lifecycle'' \cite{openai2025disrupting, ibm2025xforce, google2025advances}. In November 2025, Anthropic disclosed the first documented case of a large-scale cyberattack executed with minimal human intervention, in which a state-sponsored group manipulated Claude Code to target large financial institutions, technology companies, and government agencies \cite{anthropic2025espionage}. By 2026, generative AI-augmented malware was common among the ``industrial-scale application of generative models within adversarial workflows'' \cite{ibm2026offensive, google2026aiexploitation, verizon2026dbir}. For the purpose of this MQ, we treat ``AI-enablement'' as any use of AI by an attacker in support of a cyberattack. (AI's use in cyber defense over this period also increased. In September 2025, Anthropic reported an ``inflection point'' whereby AI models became useful for cybersecurity tasks in practice, not just theory, to help detect, analyze, and remediate vulnerabilities in code and deployed systems \cite{anthropic2025cyberdefenders}.)

\noindent\textbf{Harm estimation, $H$:} we can attempt lower-bound harm estimates using incident databases: LLM analysis of the AIID found one full match in T1, with (assessed harm count: one) and two full matches in T2 (assessed harm count 11-42). LLM analysis of the OECD AIM (pre-filtered for incidents in the financial and insurance services industry, entailing: ``economic/property'', ``reputational'', ``public interest'' or ``other'' harm) found three full matches in T1 and four in T2.

Point estimates: Tier 1: no publicly available source provides complete data on the number of AI-enabled cyberattacks matching [O] and/or [R] for any [T]. Tier 2: we can attempt to combine multiple proxy sources. Verizon's annual Data Breach Investigations Report (DBIR) draws on anonymized incident data contributed by its own incident response team, law enforcement, forensic firms, cyber insurers, and information-sharing groups worldwide. The dataset captures only reported and investigated incidents, skews toward larger organizations, and varies in contributor composition each year, so figures should be interpreted as indicative rather than comprehensive. Verizon defines an ``incident'' as a security event that compromises the integrity, confidentiality or availability of an information asset, and a ``breach'' as an incident that results in the confirmed disclosure (not just potential exposure) of data to an unauthorized party.

The 2025 DBIR (events from November 2023--October 2024; T1) analyzed 22,052 real-world security incidents across 139 countries, of which 12,195 were confirmed data breaches \cite{verizon2025dbir}. The 2026 DBIR (events from November 2024--October 2025; T2) analyzed more than 31,000 incidents across 145 countries, of which more than 22,000 were confirmed data breaches \cite{verizon2026dbir}.

Within the financial and insurance sector, the 2025 DBIR recorded 3,336 incidents and 927 confirmed breaches across organizations of all sizes, with 117 breaches attributed to large organizations (1,000 or more employees) \cite{verizon2025dbir}. The 2026 DBIR recorded 3,809 incidents and 1,300 confirmed breaches across all sizes, with 50 breaches attributed to large organizations \cite{verizon2026dbir}. The sector-level breach count increased by $\sim$40\% between T1 and T2.

The large-organization figures require careful interpretation. The breach-to-incident ratios are near unity ($\sim$0.87 in T1; $\sim$0.96 in T2), i.e., almost all reported incidents are breaches. This most likely reflects reporting selection, rather than a true conditional breach rate. Large financial institutions are subject to mandatory disclosure obligations that create strong incentives to conduct thorough forensic investigations of whether data was exfiltrated \cite{sec2024breach, morganlewis2024sec}. Incidents in which no breach occurred are likely to be resolved internally without reaching Verizon's contributing sources, engaged primarily when breaches warrant external investigation \cite{fsisac2025navigating, verizon2026dbir}. Furthermore, the sharp decline in large-organization figures between T1 and T2 (from 117 to 50 breaches) is likely to reflect noise, and year-on-year variation in Verizon's contributor composition, rather than a genuine fall in attack volume. The unknown-size category dominates the financial sector dataset in both T1 and T2 (648 and 892 unknown-size breaches respectively), and the total sector-level figures increased from 927 to 1,300 breaches. We therefore treat the large-organization figures as a noisy lower bound, rather than a point estimate.

This MQ specifies AI-enabled cyberattacks, but no available source reports financial sector breaches by degree of AI involvement. The Verizon 2026 DBIR, drawing on data from Anthropic's Safeguards Team covering 793 confirmed policy-violating ``threat actors'' between March 2025 and February 2026, found that the median actor sought AI assistance for $\sim15$ distinct techniques as per MITRE ATT\&CK (a commonly used cybersecurity knowledge base), with extreme cases reaching 40--50 techniques, representing multi-session campaigns treating Anthropic's Claude as a co-developer across the full attack chain. However, Verizon's overall assessment is that AI's primary impact is still operational: automating and scaling well-documented techniques rather than unlocking novel attack surfaces \cite{verizon2026dbir}. This characterization is consistent with the observation of the Financial Services Information Sharing and Analysis Center (FS-ISAC), an industry consortium, that attackers use LLMs primarily to lower barriers to entry and enhance existing attack vectors rather than to create novel ones \cite{fsisac2025navigating}, and with IBM X-Force's assessment that AI has changed the speed, scale, and accessibility of attacks without altering their fundamental methods \cite{ibm2026xforce}. On this basis, we assume AI-enablement, as defined in this MQ, is effectively ubiquitous in modern sophisticated cyberattacks against the financial sector.

\noindent\textbf{Trend:} \emph{Unclassifiable}. According to Verizon's data: large-organization-specific breaches \emph{decreased} between T1 and T2, but sector-level breaches increased by $\sim$40\%. With such a small proportion of large-organizations relative to total sector level figures, the data are too noisy to confirm directional change at the level of [S].

\noindent\textbf{Confidence:} N/A. (Tier 2 proxy sources are not sufficiently comprehensive or reliable to allow more than a best guess at the trend direction).

\noindent\textbf{Exposure estimation, $E$:} exposure is defined by [S] and [O]. In order to be meaningful and practically useful it must reflect the opportunity for harm, in this case the total number of cyberattacks (successful and unsuccessful; detected and undetected) on [S]. It is likely that per-firm exposure is rising due to the AI-driven democratization of offensive capability, which lowers the time and skill required to mount sophisticated attacks \cite{ibm2026xforce, fsisac2025navigating}.  However, no available combination of sources enables estimation of this trend with any confidence.

For context, we can also consider the contribution to exposure of the number of organizations that could plausibly be targeted by AI-enabled cyberattacks. We approximate this by the global count of financial-services organizations with 1,000 or more employees. No authoritative registry exists, so we combine three overlapping sources. The Financial Stability Board's 29 globally systemically important banks \cite{fsb2024gsib} provides a core that unambiguously meet the size threshold. The Forbes Global 2000 classifies roughly 500--600 companies in its financial sector \cite{forbes2024global}, of which approximately 400 fall within banking, insurance, and capital-markets sub-industries relevant to [S]; this serves as a mid-range proxy since Global 2000 inclusion correlates with, but does not strictly enforce, the 1,000-employee threshold. Extending to large private and mutual financial institutions via industry-database filtering yields an estimated 1,500--2,500 firms globally.

\noindent\textbf{Trend:} \emph{Unclassifiable}. Although we judge that effective per-organization exposure is most likely to be \emph{increasing} [$E \uparrow$] (i.e. individual organizations experience more cyberattacks, more of which are enabled in some form by AI), proxy sources do not permit even a best guess at a quantifiable level.

\noindent\textbf{Confidence:} N/A

\begin{classbox}
\noindent\textbf{Classification: \emph{Unclassifiable}.} It is likely that the use of AI to enable cyberattacks, as well as the total number of cyberattacks, increased across T1 and T2. However, neither of these trends can be quantified, and at such low confidence it is not possible to estimate whether harm-per-exposure $\hat{H}$ increased or decreased. Therefore, per \S\ref{sec:estimate}, we abstain: current public evidence does not support a classification for AI-enabled cyberattacks on large financial services organizations. 

This MQ demonstrates the difficulty of assessing intentional harms due to the lack of available data. However, cybersecurity experts may be able to augment our methods and findings qualitatively to produce a defensible classification, and government authorities, to whom financial organizations disclose incident and breach data directly, may hold data sufficient to support a quantitative classification.

\noindent\textbf{Confidence: Unclassifiable.}
\end{classbox}

\subsection{Facial-recognition misidentification and wrongful arrest.}\label{d3}
\noindent\textbf{Monitoring question:} [S] Among US individuals arrested by state or local law enforcement [O] based on investigative leads generated using facial recognition technology (FRT), [R] how many are wrongfully arrested due to facial-recognition misidentification [T] per calendar year?

\textbf{Time period:} T1: 2024 (1 January--31 December 2024); T2: 2025 (1 January--31 December 2025).

\textbf{Harm estimation, $H$:} Lower-bound harm estimate: LLM filtering of the AIID returned six full matches in 2022 (harm count six) and six full matches in the 2024 period (harm count: 15--71). Documented wrongful-arrest cases compiled by the American Civil Liberties Union \cite{aclu2024facerec} provide a supplementary floor: two documented cases in 2022, two in 2023, and two in 2024, accumulating to 10 cases across eight US states through 2024. These figures are lower bounds: they capture only cases that reached litigation or press coverage, and as such are likely to be a significant under count.

Point estimate (Tier 2): A defensible point estimate is difficult to construct. GAO-23-105607 documents approximately 60{,}000 cumulative facial-recognition searches across seven US federal law-enforcement agencies, starting on different dates between early 2018 and mid-2019, through to April 2023 \cite{gao2023facerec}, with GAO-24-107372 extending coverage into 2024 \cite{gao2024facerec}. The search volumes of state and local police (who conduct the majority of FRT-assisted policing in the US) are not systematically inventoried. The \emph{Perpetual Line-Up} study estimated that more than 117 million US adults were enrolled in police face-recognition networks as of 2016 \cite{garvie2016perpetual}, implying sustained deployment breadth.

\noindent\textbf{Trend:} \emph{Unclassifiable}. Although the lower-bound figures suggest an increase in harm between 2022 and 2024, they are individually unrepresentative and, without a point estimate, cannot be relied upon to provide a reliable signal.

\noindent\textbf{Confidence:} \emph{N/A}. Tier 3 (expert elicitation required). Tier 2 proxy sources are insufficient to provide a reliable figure for wrongful-arrest incidence: the federal-search-volume (60{,}000 cumulative), and the documented-case numbers (single-digit annual) span four orders of magnitude. 

\textbf{Exposure estimation, $E$:} We approximate exposure by the share of US jurisdictions whose law-enforcement agencies deploy facial-recognition technology. Jurisdiction-level data on the use of facial recognition are publicly available up to 2020, when 20\% (54 of 268 cities sampled) were found to use the technology \cite{johnson2024police}. Deployment is reported to be increasing rapidly, with one commercial application seeing police search volumes double between 2023 and 2024 \cite{therecord2024clearview}. With no available post-2020 estimates, we apply three trajectories to the 2003--2020 observations. For the lower bound, we fit a linear trend to all observations (conservative, as early slow-growth years attenuate the slope). The point estimate extrapolates the straight line through the 2016 and 2020 observations. For the upper bound, we fit an exponential growth model (+16\% per year), reflecting the accelerating adoption rate observed across survey waves. Under these assumptions, facial recognition deployment is estimated at 24\% (19--27\%) of jurisdictions in 2022, 25\% (19--31\%) in 2023, 27\% (20--41\%) in 2024, and 32\% (25--56\%) in 2025.

\textbf{Trend:} \emph{Increasing} [$E\uparrow$]. Deployment share is rising; the rate of increase is uncertain and spans a linear-to-exponential range.

\textbf{Confidence:} \emph{Low}. Tier 2-3 (derived from reasonable sources, but post-2020 estimates are extrapolations).

\begin{classbox}
\textbf{Classification: \textit{Unclassifiable}}. Although lower-bound counts suggest that harm increases from T1 to T2 (six to 15--71), there is insufficient data to reliably quantify any point estimates. Jurisdiction-level facial recognition deployment is rising, but it is not possible to determine the rate of increase in exposure relative to harm. Expert elicitation and/or a targeted disclosure mandate may increase confidence.
\end{classbox}

\subsection{Deepfake-enabled investment scams.}\label{d4}



\noindent\textbf{Monitoring question:} [S] Among US residents [O] exposed to online investment scams involving AI-generated deepfake media (video, image, or voice) [R] how many lose money [T] per calendar year?

\textbf{Time period:} T1: 2024 (1 January--31 December 2024); T2: 2025 (1 January--31 December 2025).

\textbf{Harm estimation, $H$:} LLM filtering of the AIID returned 13 full matches in 2024 (assessed harm count: 13--1{,}101) and 21 full matches in 2025 (assessed harm count: 23--970{,}166). 

Point estimate: Tier 2: the US Federal Bureau of Investigation (FBI) Internet Crime Complaint Center (IC3) describes itself as ``the primary destination for the public to report cyber-enabled crime and fraud as well as a key source for information on scams and cyber threats''. IC3 publishes an annual report (each April for the preceding year) that includes details of complaints received including crime types and assets lost.

AI-enabled investment fraud is a subset of the category of investment fraud. IC3 reported $\sim48$ thousand complaints of investment fraud with losses totaling \$6.57 billion in 2024 \cite{fbi2024ic3}, and $\sim73$ thousand complaints of investment fraud with losses of \$8.65 billion in 2025 (complaints increased by a factor of $\sim1.5$, and losses by a factor of $\sim1.3$). The 2025 report had, for the first time, a dedicated section on AI use in cybercrime, noting that $\sim4.4$ thousand investment fraud complaints (the losses from which totaled \$632 million) included references to AI \cite{fbi2025ic3}.

It is likely that AI is used in a higher proportion of investment frauds than this figure suggests. As noted in D.2. on AI-enabled cyberattacks on large financial services organizations, AI was assessed to be used for productivity gains across a wide range of cyberattacks from around mid-2024. Of the 2024 and 2025 investment fraud figures, respectively, around 87\% ($\sim42$ thousand complaints; \$5.8 billion in losses) and 85\% ($\sim62$ thousand complaints; \$7.28 billion in losses) are associated with a cryptocurrency-investment fraud subcategory (``pig butchering'') that the FBI associates specifically with the use of AI-generated "synthetic-media".


\noindent\textbf{Trend:} \emph{Increasing} [$H\uparrow$]. The number of investment fraud complaints and total losses increased by factors of around 1.5 and 1.3 respectively from T1 to T2. It is likely that AI use in the reported complaints also increased over the time period. The lower bound figures, although individually unrepresentative, are directionally consistent with such an increase. 

\noindent\textbf{Confidence:} \emph{Low} (Tier 2-3): The FBI IC3 is a large, mandatory-intake complaint corpus with stable methodology. However, despite being an authoritative category-adjacent proxy source, as with other intentional harms, it is difficult to know how representative any estimate based on reported data is likely to be, as many victims do not report attacks. Research quoted in 2026 by the US Federal Trade Commission (FTC) (although based on surveys carried out between 2005--2017), suggests that only around 4.8\% of victims of mass-market consumer fraud complain to a Better Business Bureau or a government entity \cite{anderson2021massmarket}. This would give harm figures of around 20 times IC3's reported complaint figures (e.g. around 850 thousand and 1.24 million victims of deepfake enabled "pig butchering" scams in 2024 and 2025 respectively, although it is also impossible to discern exactly how many of the known complaints meet the criteria for [O]). 

\textbf{Exposure estimation, $E$:}  In order to be meaningful and practically useful we need to know the total number of investment fraud solicitations (successful and unsuccessful; detected and undetected) on [S]. Unfortunately, as with other intentional harms, exposure is difficult to observe or reliably record. A plethora of reports by cybersecurity organizations suggest that the number of deepfake incidents has increased significantly since the introduction of ChatGPT in 2022: 4x increase in deepfakes detected worldwide 2023 to 2024: 1,300\% increase in deepfake fraud attempts in contact centers from 2024 to 2025; 300\% increase in face-swap attacks between 2023 and 2025 \cite{stingrai2026deepfake}. Although these figures are not directly relevant to [S] and [O], they suggest that exposure to deepfake-enabled investment fraud attacks is increasing as fast, or faster, than the successful attack rate. 

To note: whether a person falls victim to a scam or not is a complex, multi-factorial equation, and these figures tell us little about other key variables, such as the quality of each attack (although this is most likely also increasing, as AI systems become better at generating lifelike content). 

For context, we can also estimate the total population of US residents reachable by online investment frauds across email, messaging, and social-media channels. 96\% of US adults (around 249 million people) use the internet, an estimate stable since 2023 \cite{pew2024internet}.

\textbf{Trend:} \emph{increasing} [$E\uparrow$]. We judge that effective per-person exposure is most likely to be increasing [E ↑] (i.e. a given US resident will experiences more deepfake-enabled investment fraud solicitations in 2025 than in 2024). 

\textbf{Confidence:} \emph{Low} (Tier 2-3). Proxy sources are not sufficiently relevant or reliable to allow more than a best guess at the trend direction and magnitude relative to that of harm.

\begin{classbox}
\textbf{Classification: \textit{Unclassifiable.}} Harm is likely to be increasing ($H\uparrow$), exposure is also likely to be increasing ($E\uparrow$), but our estimates are not reliable enough to yield quantities or determine beyond a best guess which is rising faster. It seems more likely than not that exposure is rising faster than harm, which would give a mitigating classification, suggesting that people are learning to recognize and avoid deepfake-enabled investment scams, even though the number (and quality) of scam attempts, as well as the absolute harm caused, is increasing. Cybersecurity experts may be able to augment our methods and findings qualitatively to produce a defensible classification.
\end{classbox}

\section{Probabilistic Trajectory Classification}
\label{app:prob-classification}
Section~2.2 (Figure~4) places a monitoring question on a $2\times2$ grid from the \emph{signs} of two trends: the exposure trend and the harm-per-exposure trend. Under Tier~2 evidence those signs are uncertain, and a single hard decision both discards that uncertainty and can flip under a small perturbation. We instead carry the estimate uncertainty through the classifier, turning the grid into a \emph{distribution} over the four trajectories plus \textsc{unclassifiable}---so a borderline question reads as, e.g., $0.62$ \textsc{escalating} / $0.27$ \textsc{concentrating} rather than being forced into one cell. The procedure needs only the analyst's point estimates and uncertainty factors.

\paragraph{Model.}
Index the two time points by $t\in\{1,2\}$. We estimate four quantities: $H_t$, the number of harm events of type [R] in period $t$, and $E_t$, the number of opportunities for [R] to occur during period $t$, with $H_t,E_t>0$. Write $X$ for any one of $H_1,H_2,E_1,E_2$; the analyst supplies a point estimate $\bar X>0$ and a factor $u_X\ge 1$. Since $X>0$ and may range over orders of magnitude, we give it multiplicative scatter about $\bar X$,
\begin{equation}
X = \bar X\,e^{\sigma_X Z},\qquad Z\sim\mathcal N(0,1),
\label{eq:model}
\end{equation}
i.e.\ $X$ is log-normal with median $\bar X$. The width is fixed by the factor, $\sigma_X=\ln u_X/1.96$, so the central $95\%$ of $X$ spans $[\bar X/u_X,\,u_X\bar X]$; $u_X=2$ gives $[\bar X/2,\,2\bar X]$ and $u_X=1$ a point mass at $\bar X$. The four quantities are drawn independently.

If an incident database records $L_t$ events matching [R] in period $t$, then $H_t\ge L_t$, and we condition \eqref{eq:model} for harm on $H_t\ge L_t$. This removes the mass below $L_t$ and shifts $H_t$ appreciably only when $L_t$ is close to $\bar H_t$ (\S\ref{sec:estimate}).

\paragraph{Trends and bands.}
The two grid axes are the exposure trend $\Delta E=\ln(E_2/E_1)$ and the harm-per-exposure trend, the latter built from the ratio $\hat{H}_t=H_t/E_t$ and the harm trend $\Delta H=\ln(H_2/H_1)$:
\begin{equation}
\Delta \hat{H}=\ln\frac{\hat{H}_2}{\hat{H}_1}=\Delta H-\Delta E
=\ln\frac{H_2/H_1}{E_2/E_1}.
\label{eq:trends}
\end{equation}
($\Delta E,\Delta H,\Delta \hat{H}$ are the exposure, harm, and harm-per-exposure trends written $E,H,\hat H$ in Section~2.2.) We use the difference $\Delta H-\Delta E$ rather than a ratio of trends, which is unstable near $\Delta E=0$. Although the four quantities are sampled independently, both trends are computed from the same draws of $E_1,E_2$, which induces the negative correlation between $\Delta E$ and $\Delta\hat H$ implied by the shared denominator. A trend is \emph{flat} when it lies inside an indifference band, $|\Delta E|\le\epsilon_E$ or $|\Delta \hat{H}|\le\epsilon_{\hat H}$. A draw with neither trend flat is classified by their signs:
\begin{center}
\begin{tabular}{c|cc}
& $\Delta \hat{H}<-\epsilon_{\hat H}$ & $\Delta \hat{H}>\epsilon_{\hat H}$\\
\hline
$\Delta E>\epsilon_E$ & \textsc{mitigating} & \textsc{escalating}\\
$\Delta E<-\epsilon_E$ & \textsc{receding} & \textsc{concentrating}\\
\end{tabular}
\end{center}
If either trend is flat the draw is \textsc{unclassifiable}; with $\epsilon_E=\epsilon_{\hat H}=0$ this is exactly the deterministic rule of Figure~4.

\paragraph{Raw weights.}
The raw weight of each class is its probability under the input uncertainty, $w(c)=\Pr(C=c)$ with $\sum_c w(c)=1$, estimated by Monte~Carlo (Algorithm~\ref{alg:prob-class}): draw $N$ samples of $H_1,H_2,E_1,E_2$ from \eqref{eq:model} (truncating harm where a floor is given), classify each draw, and average the class indicators. Without truncation $(\Delta E,\Delta\hat H)$ is exactly bivariate normal and the raw weights are its band probabilities; we use Monte~Carlo so the same code also covers the truncated case. Alongside the weights we report the mean and $95\%$ interval of $\Delta H,\Delta E,\Delta\hat H$; the interval on $\Delta\hat H$ shows how firmly its sign is identified.

\paragraph{Directional-confidence adjustment.}
The raw weights count a draw as classifiable whenever both trends clear their bands, regardless of whether the \emph{sign} of each trend is identified across draws: if $\Delta E$ is positive in $51\%$ of draws and negative in $49\%$, almost every draw clears a small band, and the near-total ignorance about the direction of exposure surfaces only as a split between quadrants rather than as \textsc{unclassifiable} mass. We therefore cap the classifiable mass by the directional confidence of each axis,
\begin{equation}
c_E=\bigl|\Pr(\Delta E>\epsilon_E)-\Pr(\Delta E<-\epsilon_E)\bigr|,
\qquad
c_{\hat H}=\bigl|\Pr(\Delta \hat H>\epsilon_{\hat H})-\Pr(\Delta \hat H<-\epsilon_{\hat H})\bigr|,
\label{eq:confidence}
\end{equation}
estimated by empirical frequencies over the $N$ draws; each $c$ is $1$ when the sign is certain and $0$ at a coin flip. Treating the axes as independent directional evidence gives classifiable mass $\pi=c_E\,c_{\hat H}$. Writing $\mathcal T$ for the set of four trajectories, the reported weights keep the raw proportions within $\mathcal T$ but rescale their total to $\pi$:
\begin{equation}
\tilde w(c)=\pi\,\frac{w(c)}{\sum_{c'\in\mathcal T} w(c')}
\quad\text{for } c\in\mathcal T,
\qquad
\tilde w(\textsc{unclassifiable})=1-\pi.
\label{eq:adjusted}
\end{equation}
When both signs are certain ($c_E=c_{\hat H}=1$) the adjustment is the identity, so setting every $u_X=1$ and $\epsilon_E=\epsilon_{\hat H}=0$ still recovers the deterministic grid of Figure~4 exactly. All classification probabilities reported in this paper are the adjusted weights $\tilde w$; the implementation also returns the raw weights $w$ for diagnostic use.

\begin{algorithm}[t]
\caption{Probabilistic trajectory classification}
\label{alg:prob-class}
\begin{algorithmic}[1]
\REQUIRE point estimates $\bar H_1,\bar H_2,\bar E_1,\bar E_2$; factors $u_\cdot$; optional floors $L_1,L_2$; bands $\epsilon_E,\epsilon_{\hat H}$; draws $N$
\ENSURE adjusted weights $\tilde w(\cdot)$; raw weights $w(\cdot)$; summaries of $\Delta H,\Delta E,\Delta\hat H$
\STATE set $\sigma_X=\ln u_X/1.96$ for each $X$ (Eq.~\eqref{eq:model})
\FOR{$k=1$ to $N$}
  \STATE draw $H_1,H_2$ (truncated at $L_1,L_2$ if given) and $E_1,E_2$ from \eqref{eq:model}
  \STATE $\Delta E\leftarrow\ln(E_2/E_1)$;\quad $\Delta \hat H\leftarrow\ln(H_2/H_1)-\Delta E$
  \STATE assign $C^{(k)}$ from $(\Delta E,\Delta \hat H)$ by the banded rule
\ENDFOR
\STATE $w(c)\leftarrow\tfrac{1}{N}\sum_{k}\mathbf 1[C^{(k)}=c]$ for each class
\STATE estimate $c_E,c_{\hat H}$ from the draw frequencies (Eq.~\eqref{eq:confidence}); $\pi\leftarrow c_E\,c_{\hat H}$
\STATE $\tilde w(\cdot)\leftarrow$ renormalize the four trajectory weights to total $\pi$; $\tilde w(\textsc{unclassifiable})\leftarrow 1-\pi$ (Eq.~\eqref{eq:adjusted})
\end{algorithmic}
\end{algorithm}

\end{document}